\documentclass[fleqn,usenatbib]{mnras}

\usepackage{newtxtext,newtxmath}

\usepackage[T1]{fontenc}
\usepackage{ae,aecompl}

\usepackage{graphicx}	
\usepackage{amsmath}	
\usepackage{amssymb}	
\usepackage{hyperref}	
\usepackage{pdflscape}	

\newcommand\Spitzer{$\it Spitzer$} 
\newcommand\Herschel{$\it Herschel$} 
\newcommand\HII{\mbox{H\,\scriptsize{II }}} 
\newcommand\UCHII{\mbox{UC H\,\scriptsize{II }}} 
\newcommand{\kms}{\mbox{km\,s$^{-1}$}} 

\title[Star formation in IRDC G31.97+0.07]{Star formation in IRDC G31.97+0.07}

\author[Zhou et al.]{Chenlin Zhou,$^{1,2,3}$ Ming Zhu,$^{1,3}$\thanks{Email: mz@nao.cas.cn} Jinghua Yuan,$^1$ Yuefang Wu,$^4$ Lixia Yuan,$^{1,2,3}$, \newauthor T. J. T. Moore$^5$, D. J. Eden$^5$
\\
$^{1}$ National Astronomical Observatories, Chinese Academy of Sciences, 20A Datun Road, Chaoyang District, Beijing 100012, China
\\
$^{2}$ University of Chinese Academy of Sciences, Beijing 100049, China
\\
$^{3}$ Key Laboratory of FAST, NAOC, Chinese Academy of Science, Beijing 100012, China
\\
$^{4}$ Department of Astronomy, Peking University, 100871 Beijing, China
\\
$^{5}$Astrophysics Research Institute, Liverpool John Moores University, IC2, Liverpool Science Park, 146 Brownlow Hill, Liverpool, L3 5RF, UK
}

\date{Accepted XXX. Received YYY; in original form ZZZ}

\pubyear{2018}

\begin{document}
\label{firstpage}
\pagerange{\pageref{firstpage}--\pageref{lastpage}}
\maketitle



\begin{abstract}
We utilize multiple-waveband continuum and molecular-line data of CO isotopes, to study the dynamical structure and physical properties of the IRDC G31.97+0.07. 
We derive the dust temperature and H$_2$ column density maps of the whole structure by SED fitting. The total mass is about $2.5\times10^5\,M_{\sun}$ for the whole filamentary structure and about $7.8\times10^4\,M_{\sun}$ for the IRDC.
Column density PDFs produced from the column density map are generally in the power-law form suggesting that this part is mainly gravity-dominant. The flatter slope of the PDF of the IRDC implies that it might be compressed by an adjacent, larger \HII region.
There are 27 clumps identified from the 850\,\micron \ continuum located in this filamentary structure. 
Based on the average spacing of the fragments in the IRDC, we estimate the age of the IRDC. 
The age is about $6.4\,$Myr assuming inclination angle $i = 30\degr$.
For 18 clumps with relatively strong CO and $^{13}$CO (3-2) emission, we study their line profiles and stabilities.
We find 5 clumps with blue profiles which indicate gas infall motion and 2 clumps with red profiles which indicate outflows or expansion. 
Only one clump has $\alpha_\mathrm{vir} > 2$, suggesting that most clumps are gravitationally bound and tend to collapse.
In the Mass-$R_\mathrm{eq}$ diagram, 23 of 27 clumps are above the threshold for high-mass star formation, suggesting that this region can be a good place for studying high-mass star-forming.
\end{abstract}

\begin{keywords}
stars: formation -- ISM: clouds -- ISM: kinematics and dynamics -- ISM: structure
\end{keywords}




\section{Introduction}
Massive stars ($M \geq 8 M_{\sun}$) and massive stellar clusters play an important role in moulding the galactic environment and determining the metal enrichment through their ionizing radiation, stellar winds, outflows and explosive deaths.
However, much remains to be understood about the formation and protostellar evolution of high-mass stars.
The initial conditions and the earliest evolutionary stages of massive star formation are still unclear.

Infrared dark clouds (IRDCs) have been proposed to be good candidates for the birthplaces of massive stars and clusters. IRDCs were first observed as dark silhouettes against the bright Galactic background infrared (IR) emission by the {\it Midcourse Space Experiment} (MSX; \citealt{Carey1998ApJ,Egan1998ApJ,Simon2006ApJ}), and the {\it Infrared Space Observatory} \citep{Hennebelle2001A&A}.
Previous molecular lines and continuum studies suggested that IRDCs were cold ($<20\,$K) and dense ($>10^5\,\mathrm{cm}^{-3}$) gas collections, typically arranged in filamentary and/or globular structures with compact cores. Their masses range from $10^2$ to $10^5\,M_{\sun}$ with a scale of several to tens of pc \citep{Carey1998ApJ,Egan1998ApJ,Rathborne2006}. All of these properties imply that IRDCs are an excellent mass reservoir for massive star formation.

The filamentary IRDC can be seen as a self-gravitating gas cylinder. Theoretical work predicted that the filament should fragment into multiple cores with quasi-regular spacing due to the ``sausage" instability \citep{Chandrasekhar1953ApJ,Nagasawa1987,Inutsuka1992,Tomisaka1995}.
That was approved by observational studies showing that many cores/clumps were embedded in filamentary IRDCs \citep{Rathborne2006,Wang2008ApJ,Henning2010A&A,Jackson2010ApJ,Wang2014MNRAS}.
In addition, recent studies have revealed massive young stellar objects (YSOs) and cores in IRDCs \citep{Rathborne2006,Rathborne2007ApJ,Rathborne2011ApJ,Beuther2007ApJ,Wang2008ApJ,Henning2010A&A},
indicating that IRDCs are active subjects with undergoing massive star formation.

G31.97+0.07 was first identified by {\it MSX} \citep{Simon2006ApJ}. The IR morphology has a long, thin filamentary structure, and it is located in the west side of an active star-forming complex containing several (UC)\HII regions (see Figure~\ref{fig:RGB} in Sec.~\ref{sec:continuum}).
The central velocity of the IRDC is $96.7\,$\,\kms \ given by \citet{Simon2006ApJ}. 
The central velocity and position are well consistent with the molecular cloud GRSMC G032.09+00.09 \citep{Roman-Duval2009}.
So in this work, we use the distance of GRSMC G032.09+00.09 for our object. 
The distance is $7.07\,$kpc, using the Clemens rotation curve of the Milky Way \citep{Clemens1985ApJ} and solving the kinematic distance ambiguity by H\,{\scriptsize{I}} self-absorption analysis. 

\citet{Rathborne2006} observed nine dense clumps with masses ranging from $151 \ \mathrm{to} \ 1890\,M_{\sun}$ using the IRAM 30\,m single-dish telescope.
\citet{Wang2006ApJ} reported water masers in G31.97+0.07, and \citet{Urquhart2009A&A} identified an \HII region in G31.97+0.07.
\citet{Battersby2014a} studied G31.97+0.07 in the NH$_3 \ (1,\ 1), \ (2,\ 2) \mathrm{and} \ (4,4)$ transitions with the VLA. They identified 11 dense cores less 0.1\,pc in size, and found that those cores were virially  unstable to gravitational collapse. They also reported that turbulence likely set the fragmentation length scale in the filament. Moreover, they note the existence of at least three large bubbles around the filament, which are likely older \HII regions. They suggested that those large \HII regions may have compressed the molecular gas to form and/or shape the IRDC and trigger more recent massive star formation.
Using the IRAM 30\,m and CSO 10.4\,m telescopes, \citet{Zhang2017A&A} performed observations with HCO$^+$, HCN, N$_2$H$^+$, C$^{18}$O, DCO$^+$, SiO and DCN toward G31.97+0.07. They reported that C$^{18}$O emission may be heavily depleted at the peak positions of some cold cores. And they suggest that some active cores are collapsing while their envelopes are expanding.

In this work, we utilize multiple wavebands of continuum data (from Mid-IR to 850\,\micron), and molecular line data of three CO isotopes at $J=1 \rightarrow 0$ and $J=3 \rightarrow 2$ transition, to study the dynamical structure of IRDC G31.97+0.07 and star formation processes in dense clumps embedded in the structure. 
The data used in this work are described in Sec.~\ref{sec:Data}.
We present results of large scale cloud and the dust and gas properties derived from continuum and molecular lines data in Sec.~\ref{sec:Result}.
Further discussion about gas dynamics and dense clumps properties is given in Sec.~\ref{sec:Discussion}.
We summarize our work in Sec.~\ref{sec:Summary}.

\section{Observations and Data Reduction}
\label{sec:Data}
\subsection{IR to Sub-millimetre Continuum Data}
\label{sec:continuum}

We utilized near-IR and mid-IR data from the GLIMPSE and MIPSGAL survey. 
The GLIMPSE survey observed the Galactic plane with four IR bands (3.6, 4.5, 5.8, 8.0\,\micron) of the IRAC instrument on the \Spitzer \ Space Telescope, and the sky coverage is $|b|<1\degr$ for $10\degr < \ell < 65\degr$ \citep{Benjamin2003,Churchwell2009PASP}.
MIPSGAL is a survey of the same area as GLIMPSE, using MIPS instrument at 24 and 70\,\micron \ on \Spitzer\ 
\citep{Carey2009}.
The point-source catalogue from GLIMPSE \citep{GLIMPSE2009} and 24\,\micron \ point-source catalogue from MIPSGAL \citep{Gutermuth2015} have been used to identify young stellar object (YSO) candidates .

Far-IR data were obtained from the Hi-GAL survey to investigate dust properties of our object. Hi-GAL (\Herschel \ Infrared Galactic Plane survey, \citealt{Molinari2010PASP}) is a key project of \Herschel \ Space Observatory which mapped the entire Galactic plane with $|b| \leq 1\degr$ The Hi-GAL data were observed in the fast mode using the PACS (70 and 160 \,\micron) and SPIRE (250, 350, 500\,\micron) instruments in the parallel mode. We used the \Herschel \ Interactive Processing Environment (HIPE) to download the continuum maps from the \Herschel \ Science Archive and to reduce the data using standard pipelines. The effective angular resolution is $10.\!\!\arcsec2, \ 13.\!\!\arcsec5, 18.\!\!\arcsec1, \ 24.\!\!\arcsec9, \  \mathrm{and} \ 36.\!\!\arcsec4$ at 70, 160, 250, 350, and 500\,\micron, respectively. 
In addition, the 70\,\micron \ point-source catalogues from Hi-GAL \citep{Molinari2016} were used to investigate evolutionary stages of the clumps in our object.

We extracted 850\,\micron \ continuum data from the James Clerk Maxwell Telescope (JCMT) Plane Survey (JPS) \citep{Moore2015,Eden2017}. JPS is a targeted, yet unbiased, survey of the inner Galactic Plane in the longitude range $7\degr < \ell <63\degr$ and latitude range $|b|<0.8\degr$ using Sub-millimetre Common-User Bolometer Array 2 (SCUBA-2; \citealt{Holland2013}) at 850\,\micron \ with an angular resolution of 14.5\arcsec. 
The survey observing strategy is to map six large, regularly spaced fields within the area. Each individual survey field is sampled using a regular grid of eleven circular tiles with a uniform diameter of one degree, observed using the {\it pong}3600 mode \citep{Bintley2014}.
The data reduction process removes structures larger than 480 arcsec from the final maps, which are gridded to 3-arcsec per pixel. The average smoothed pixel-to-pixel noise is $7.19\,\mathrm{mJy\,beam}^{-1}$.
The JPS 850\,\micron \ compact-source catalogue identified by \citet{Eden2017} using the {\sc Fellwalker} algorithm \citep{Berry2015}  is also adopted in this study.

\subsection{Molecular Line Data}
\label{sec:line data}
$^{13}$CO (1-0) emission data from the Boston University-Five College Radio Astronomy Observatory Galactic Ring Survey (BU-FCRAO GRS; \citealt{Jackson2006}) were used to identify large-scale structures.
The survey covered a longitude range of $18\degr<\ell<55.\!\!\degr7$ and a latitude range of $|b|<1\degr$, using the SEQUOIA multipixel array on the Five College Radio Astronomy Observatory 14\,m telescope.  
The velocity resolution of the survey is 0.21\,\kms , and angular resolution is 46\arcsec. The typical rms sensitivity is $\sigma (T_A^*) \sim 0.13 \mathrm{K}$.

We used CO (3-2) data from the CO High Resolution Survey (COHRS; \citealt{Dempsey2013ApJS}) to study warm and high-velocity gas that may be excited by shocks and outflows from active star-forming regions.
This survey covers $|b|<0.5\degr$ between $10\degr <\ell < 65\degr$ using the HARP instrument on the JCMT.
The final data cube were smoothed to the angular resolution of 16.6 arcsec and rebinned to 1\,\kms. The mean rms is about 1\,K and a main-beam efficiency of $\eta_\mathrm{mb} = 0.61$.

Data from the $^{13}$CO/C$^{18}$O (3-2) Heterodyne Inner Milky-Way Plane Survey (CHIMPS: \citealp{Rigby2016}) were used to investigate the denser part (e.g. clumps embedded in the structure). CHIMPS was carried out using the HARP on the JCMT, observing $^{13}$CO and C$^{18}$O (3-2) simultaneously. The survey covers $|b| \leq 0.5 \degr$ and $28 \degr \lesssim \ell \lesssim 46 \degr$, with an angular resolution of 15 arcsec in 0.5\,\kms. The mean rms of the data is about 0.6\,K, and a main-beam efficiency of $\eta_\mathrm{mb} = 0.72$.


\section{Results}
\label{sec:Result}
\subsection{IR and Sub-mm Continuum Emission}
	\label{sub:Continuum Emission}

\begin{figure*}
	\centering
	\includegraphics[width=\textwidth,trim=0 0 0 0,clip]{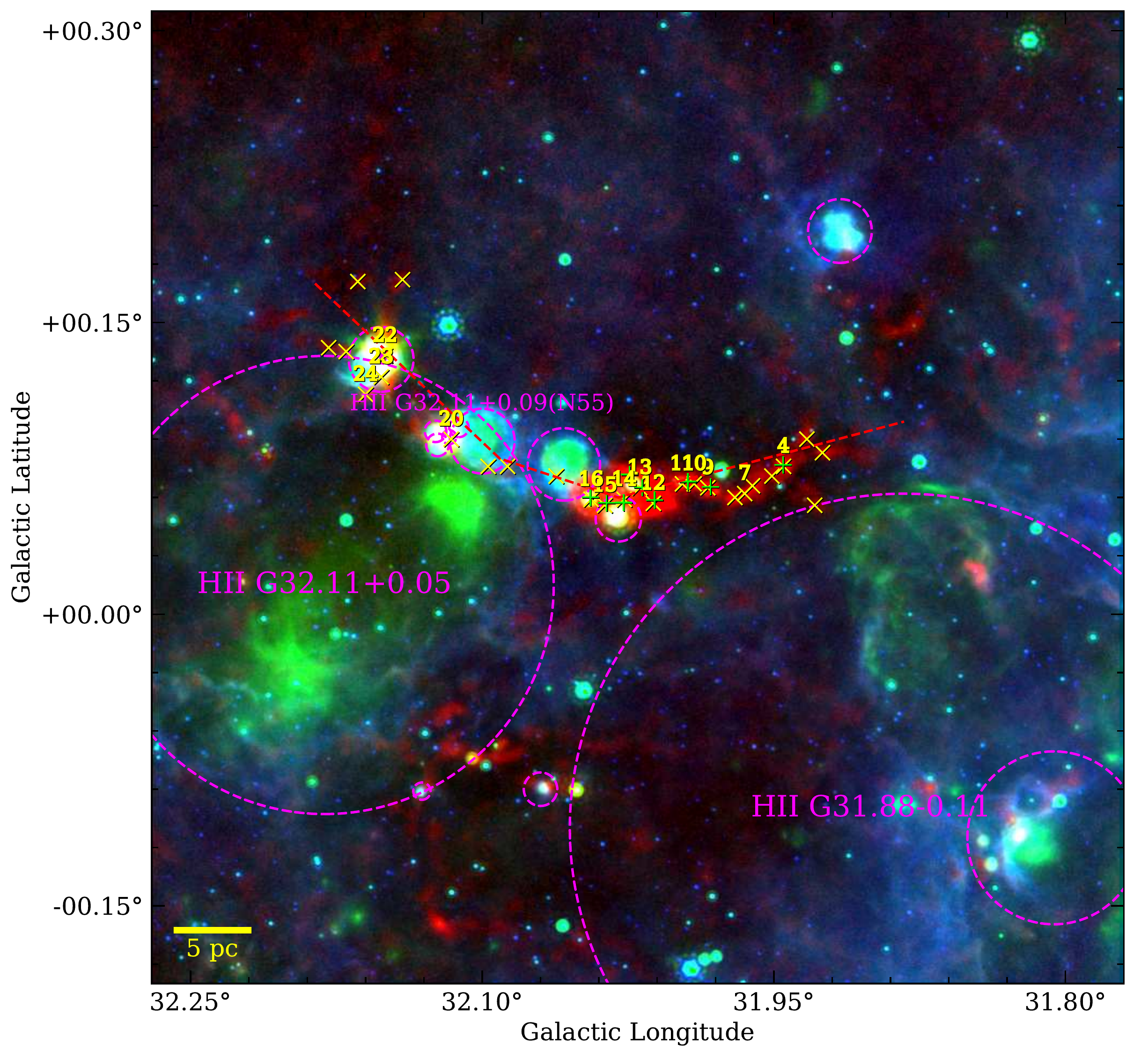}
    \caption{The three-colour image for IRDC G31.97+0.07 at large scale, red: JPS 850\,\micron, green: MIPSGAL 24\,\micron, blue: GLIMPSE 8\,\micron. Red dashed lines mark the filamentary structure. The IRDC displays the extinction feature at near/mid-IR wavebands but emits at sub-millimetre wavelengths. The magenta dashed circles represent the positions of \HII regions identified by \citet{Anderson2014}. Green crosses show the positions of the millimetre cores MM1-MM9 identified by \citet{Rathborne2006} and yellow crosses show the positions of 27 JPS compact sources identified by \citet{Eden2017}. The IDs of JPS clumps with signal-noise ratio at 850\,\micron \ larger than 10 are labelled}
    \label{fig:RGB}
\end{figure*}

Figure~\ref{fig:RGB} displays the three-colour image (red: JPS 850\,\micron, green: MIPSGAL 24\,\micron, blue: GLIMPSE 8\,\micron) of the filamentary IRDC G31.97+0.07 and its adjacent context. The filamentary structure is marked by the red-dashed line, and \HII regions identified by \citet{Anderson2014} from WISE data are shown in magenta dashed circles. In Figure~\ref{fig:RGB}, the filamentary structure is divided into two parts: the eastern near/mid-IR dominant part and western sub-millimetre dominant part. The bright 8\,\micron \ emission is mainly attributed to polycyclic aromatic hydrocarbons (PAHs) excited by the UV radiation in the adjacent \HII region \citep{Pomar2009}. Thus, 8\,\micron \ emission can be utilized to delineate infrared bubbles. The eastern part, located on the rim of the large \HII region G32.11+0.05, contains several 8\,\micron-bright \HII regions: \HII G32.16+0.13, \HII G32.11+0.09 (or bubble N55 in \citealt{Churchwell2006}), and \HII G32.06+0.08. This indicates that the eastern part hosts some young O or B type stars with strong stellar winds. Two \UCHII regions located on the rim of N55 may suggest efficient star-forming in this region. The western part, mainly consisting of IRDC G31.97+0.07 and some other small filaments, displays a dark extinction feature. The green crosses show the positions of 9 cores, MM1-MM9, identified in 1.2\,mm continuum data by \citet{Rathborne2006} using the IRAM 30\,m telescope. The yellow crosses show the positions of 27 JCMT compact sources identified by \citet{Eden2017} from JPS 850\,\micron \ data.

The 24\,\micron \ continuum emission is mainly contributed by the thermal radiation of warm dust heated by protostars \citep{Battersby2011,Guzman2015}. \HII regions and infrared bubbles, distributed in the filamentary structure, exhibit strong emission in their inner regions in the 24\,\micron \ waveband. IRDC 31.97+0.07 mostly appears as dark extinction at 24\,\micron \ and 70\,\micron. Some of the mm continuum cores identified by \citet{Rathborne2006} and sub-mm compact sources identified by \citet{Eden2017} do not show 24\,\micron \ emission, indicating that they are cold and dense. \citet{Battersby2014a} divided the IRDC into two parts, the `active part' and the `quiescent part'. MM1, MM2, MM3, MM6 and MM9, belonging to the `active part', are associated with mid-IR bright point sources. This suggests that these cores are undergoing star formation and heating the nearby dust.

Sub-mm continuum emission can be a good tracer of the cold and dense dust. It also can be an effective tool to search for clumps lacking signs of the star-forming process, or in early evolutionary stage. The long filament of IRDC G31.97+0.07 can be easily distinguished in the 850\,\micron \ image. The Bubble N55 and \HII G32.06+0.08 are devoid of 850\,\micron \ emission, while other \HII regions in the IRDC are associated with bright 850\,\micron \ clumps. This indicates that bubble N55 and \HII G32.06+0.08 are more evolved than other \HII regions in the filament. Considering the morphology in the continuum emission, we suggest that IRDC 31.97+0.07 and the \HII regions on its east side may be one continuous structure, which is further confirmed by the continuity of velocity (see in Sec.~\ref{sec:CO Emission}).


\subsection{CO Molecular Line Emission}
	\label{sec:CO Emission}
    
\begin{figure*}
	\centering
    \includegraphics[width=\textwidth,trim=0 50 50 70,clip]{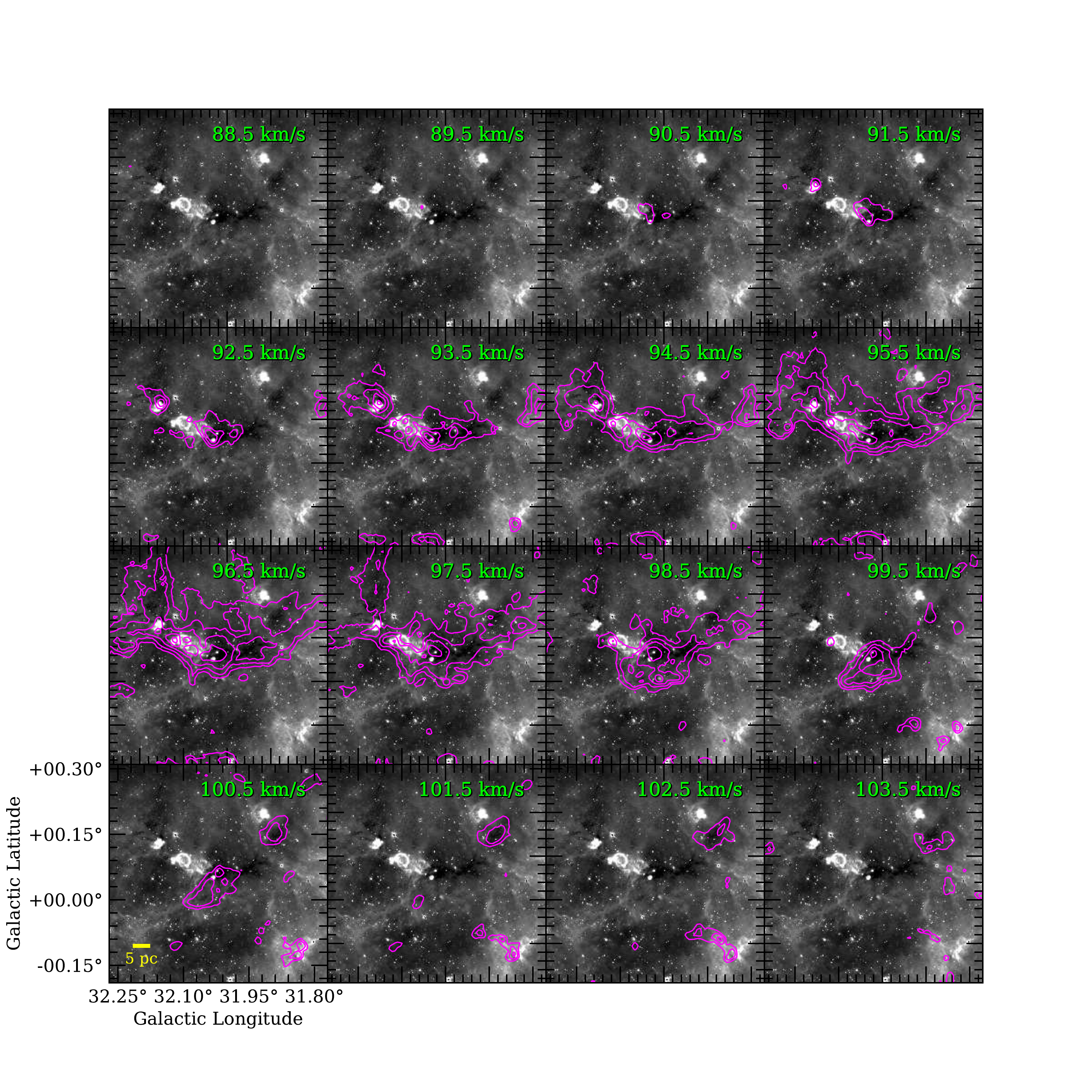}
    \caption{GRS $^{13}$CO (1-0) channel maps in steps of 1\,\kms. The background is \Spitzer-IRAC 8\,\micron \ continuum emission. Central velocities are shown in each map.}
    \label{fig:channel map}
\end{figure*}

We use GRS $^{13}$CO (1-0) molecular line emission to show the gas distribution of IRDC G31.97+0.07. Figure~\ref{fig:channel map} shows the $^{13}$CO channel maps overlaid on 8\,\micron \ emission in steps of 1\,\kms, between 90 and 101\,\kms. Most $^{13}$CO emission is detected in the velocity interval of 93 to 99\,\kms, while \UCHII regions and the `active' clumps in the IRDC contain high-velocity components which may be driven by outflows from YSOs.
The channel maps also show that IRDC G31.97+0.07 and the \HII regions on its east side belong to the continuous filamentary structure also implied by continuum emission data, with a systemic velocity at 96.5\,\kms. 

\begin{figure*}
	\centering
    \includegraphics[width=0.9\textwidth,trim=10 5 10 20,clip]{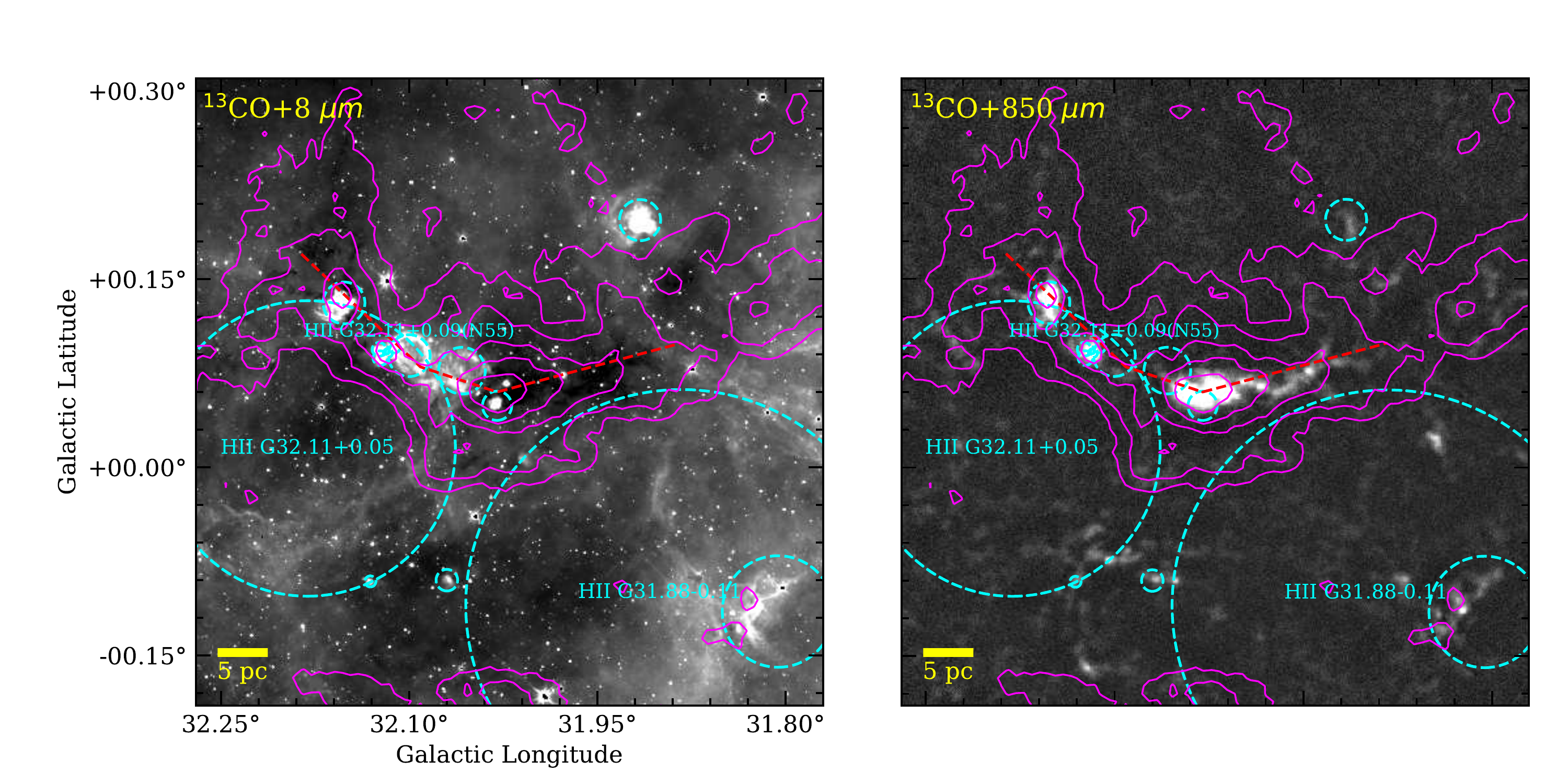}
    \caption{Contours of the GRS $^{13}$CO (1-0) integrated map overlaid on \Spitzer \ 8\,\micron \ emission map(left) and JCMT 850\,\micron \ emission map (right). The integrated velocity is from 90 to 101\,\kms. The contour levels are $[ 0.1,0.2,0.35,0.5,0.75 ] \times 31.9$\,K\,\kms. The red dashed line marks the filamentary structure. Cyan dashed circles represent the positions of \HII regions identified by \citet{Anderson2014}.}
    \label{fig:contour}
\end{figure*}

Figure~\ref{fig:contour} shows contours of GRS $^{13}$CO (1-0) integrated intensity overlaid on 8\,\micron \ and 850\,\micron \ continuum emission. The velocity interval is between 90 and 101\,\kms, with central velocity 96.5\,\kms. The red dashed line marks the filamentary structure. The $^{13}$CO (1-0) integrated map shows the molecular cloud GRSMC G032.09+00.09 identified by \citet{Rathborne2009} with central velocity 96.85\,\kms \ \citep{Roman-Duval2009}. The shape of the dense and gas-rich part of GRSMC G032.09+00.09 is consistent with the filamentary structure of the continuum. That suggests that IRDC 31.97+0.07 and the \HII regions on its east side are one continuous structure embedded in the molecular cloud GRSMC G032.09+00.09, and the distance of the cloud is $7.07$\,kpc \citep{Roman-Duval2009}.
The length of the whole filamentary structure is about 20\,arcmin, and $\sim 41$\,pc at this distance.
The radii of the JPS compact sources located in the structure range from 0.2 to 0.8\,pc, hence they are molecular clumps.

\subsection{Spectral Energy Distribution (SED) Fitting}
	\label{sec:SED}
  
To determine the properties of the dust, we fit the SED of our object pixel by pixel based on \Herschel \ data. As the most part of the IRDC is 70\,\micron \ dark, we just take emission at 160, 250, 350, 500\,\micron \ into account in our SED fitting.

In order to measure the flux of multi-waveband observations in different angular resolutions, we convolved our image to a common angular resolution by a Gaussian kernel with FWHM equal to $\sqrt{\theta_{500}^2-\theta_{\lambda}^2}$, where $\theta_{500}$ is FWHM beam size of Hi-GAL 500\,\micron \ band and $\theta_{\lambda}$ is the FWHM beam size of each Hi-GAL band. Then we re-gridded our data to be aligned pixel-by-pixel with the same pixel size of 14\arcsec. 

We used the smoothed data to fit the SED for each pixel by using a grey-body model:
	\begin{equation}
		I_\nu = B_\nu(T_d)(1-e^{-\tau_\nu}),
	\end{equation}
where $I_\nu$ is the monochromatic intensity, $B_\nu(T_d)$ is the Planck function and $\tau_\nu$, the optical depth at frequency $\nu$, is given by
	\begin{equation}
		\tau_\nu = \mu_{\mathrm{H}_2} m_\mathrm{H} \kappa_\nu N_{\mathrm{H}_2} / \mathrm{GDR}.
	\end{equation}
Here $\mu_{\mathrm{H}_2}=2.8$ is the mean molecular weight of molecular hydrogen \citep{Kauffmann2008}, $m_H$ is the mass of a hydrogen atom, $N_{\mathrm{H}_2}$ is the column density of $H_2$, GDR is the gas-to-dust ratio by mass and is assumed to be 100. The dust opacity $\kappa_\nu$ can be approximated by a power law,
	\begin{equation}
		\kappa_\nu = \kappa_{\nu_0} \left(\dfrac{\nu}{\nu_0}\right)^{\beta},
	\end{equation}
where $\nu_0=600\,\mathrm{GHz}$, $\beta$ fixed to be $2.0$ is the dust emissivity index and $\kappa_{\nu_0}=3.33\,\mathrm{cm}^2 \,\mathrm{g}^{-1}$ (from column 5 of table 1 in \citealt{Ossenkopf&Henning1994}), but scaled down by a factor of 1.5 as suggested in \citet{Kauffmann2010ApJa}.

The fitting is performed by using {\sc curve\_fit} in the Python package {\sc scipy.optimization}\footnote{\href{https://docs.scipy.org/doc/scipy/reference/tutorial/optimize.html}{https://docs.scipy.org/doc/scipy/reference/tutorial/optimize.html}}. 
For pixels with intensity in 160\,\micron \ lower than $60\,\mathrm{MJy}\,\mathrm{sr}^{-1}$ (about $3\sigma$), we found that the fitting result was not so reliable. So, for those pixels, only 250, 350, 500\,\micron \ data were used in fitting.

\begin{figure*}
	\centering
	\includegraphics[width=0.95\textwidth,trim=0 20 0 40,clip]{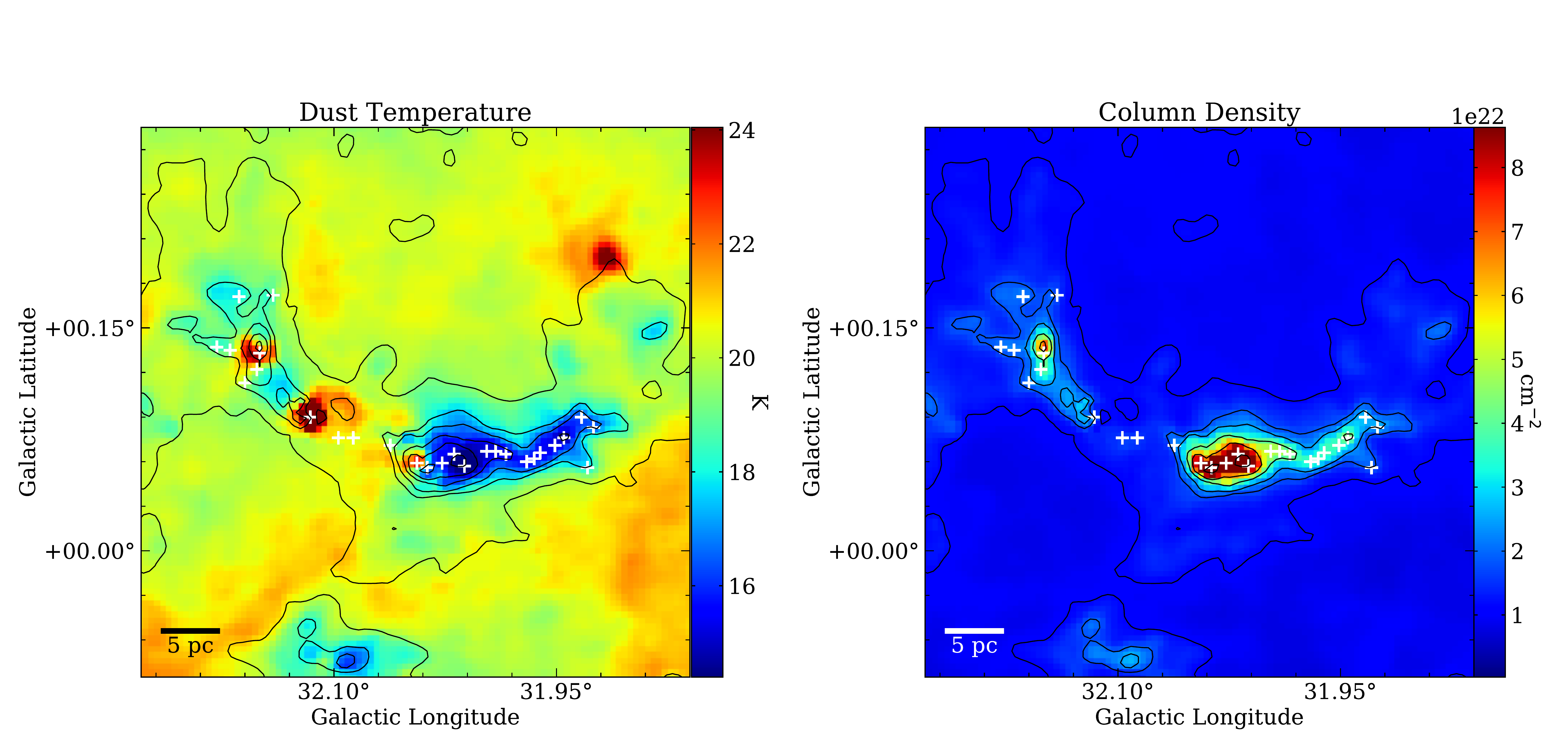}
    \caption{The dust temperature (left) and the $\mathrm{H}_2$ column density (right) maps produced by SED fitting. White crosses show the JPS compact continuum sources. The contour levels are $[ 1.1, 1.8, 2.4, 4.0, 6.5, 11.5 ] \times 10^{22}$\,cm$^{-2}$.}
    \label{fig:SED result}
\end{figure*}

The dust temperature and column density maps produced by SED fitting are presented in Figure~\ref{fig:SED result}. The dust temperature of the whole filamentary structure ranges from $10$ to $30$\,K, and the column density ranges from $0.5 \times 10^{22}$ to $10^{23}\,\mathrm{cm}^{-2}$. 
Most of the IRDC is cold (lower than $18$\,K) with relatively high column density (above $2 \times 10^{22}\,\mathrm{cm}^{-2}$), while the \HII and \UCHII regions have higher dust temperature and lower column density. 

We find that the contour of $1.1 \times 10^{22}$\,cm$^{-2}$ is similar to the CO integrated map of the cloud, and the contour of $1.8 \times 10^{22}$\,cm$^{-2}$ is consistent with the morphology of the IRDC G31.97+0.07 in IR maps. 
So, choosing those values as thresholds, we derive that the mass of the whole structure is $\sim 2.5 \times 10^5\,M_{\sun}$ and the mass of the IRDC is $\sim 7.8 \times 10^4\,M_{\sun}$.


The clump mass and dust temperature can be directly derived from the SED fitting result: 
\begin{equation}
	M_\mathrm{clump} = \mu_\mathrm{H_2} m_\mathrm{H} d^2 \Omega_\mathrm{pix} \sum N_\mathrm{H_2} ,
\end{equation}
here, $d=7.07\,\mathrm{kpc}$ is the distance of the filament structure, $\Omega_\mathrm{pix}$ is solid angular size of one pixel and $\sum N_\mathrm{H_2}$ is the sum of the column density value for all the pixels in the clump.

The source average H$_2$ column densities are calculated as:
\begin{equation}
	N_{\mathrm{H}_2} = \sum N_\mathrm{H_2} / \mathrm{N}_\mathrm{sour},
\end{equation}
where $\mathrm{N}_\mathrm{sour}$ is the number of pixels located in the clump.
The average H$_2$ volume densities are derived via:
\begin{equation}
	n_{\mathrm{H}_2} = \frac{M_\mathrm{clump}}{(4/3) \pi r_{\rm eq}^3 \,\mu_{\mathrm{H}_2} \, m_\mathrm{H}},
\end{equation}
and the mass surface densities are given by:
\begin{equation}
	\Sigma_\mathrm{mass} = \frac{M_\mathrm{clump}}{\pi r_{\rm eq}^2},
\end{equation}
where $r_{\rm eq}$ is equivalent radius.

We also determined the integrated intensity in the SED fitting process, so luminosities of sources are derived from:
\begin{equation}
	L_\mathrm{clump} = 4 \pi d^{2} \Omega_\mathrm{pix} \sum I_\mathrm{int},
\end{equation}
where $\sum I_\mathrm{int}$ is the sum of the frequency-integrated intensity($I_\mathrm{int}=\int I_\nu\,d\nu$) for each pixel in the clump.
And the luminosity-mass ratio is given by:
\begin{equation}
	\mathrm{L}/\mathrm{M} \ \mathrm{ratio} = L_\mathrm{clump}/M_\mathrm{clump}
\end{equation}

Table~\ref{tab:photometry} lists the resulting dust temperature, column density, volume density, mass surface density, mass, luminosity and luminosity-mass ratio of each source.

\begin{landscape}
\begin{table}
\centering
\caption{
Physical properties of JPS compact sources derived from SED fitting. The columns are as follows: (1)-(6): JPS source name, coordinates information and source size come from \citet{Eden2017}; (7) equivalent radius, calculated by: $d \sqrt{\mathrm{A} / \pi}$, where $d = 7.07\,\mathrm{kpc}$ and A is the area of the source; (8)-(14): dust temperature, column density, mass surface density, mass, luminosity and luminosity-mass ratio obtained from the SED fitting result.
}
\label{tab:photometry}
\begin{tabular}{ccccccccccccccc}
\hline
ID & Name & $\ell_{\mathrm{peak}}$ & $b_{\mathrm{peak}}$ & maj & min & PA & $\mathrm{r}_\mathrm{eq}$ & $\mathrm{T}_\mathrm{dust}$ & $\mathrm{N}_{\mathrm{H}_2}$ & $\mathrm{n}_{\mathrm{H}_2}$ & $\Sigma_\mathrm{mass}$ & Mass & Luminosity & L/M ratio \\
 & &  &  & \arcsec & \arcsec & \degr & pc & K & $10^{22}\,\mathrm{cm}^{-2}$ & $10^4\,\mathrm{cm}^{-3}$ & $\mathrm{g}\,\mathrm{cm}^{-2}$ & $\mathrm{M}_{\sun}$ & $\mathrm{L}_{\sun}$ & $\mathrm{L}_{\sun} / \mathrm{M}_{\sun}$ \\
 & (1) & (2) & (3) & (4) & (5) & (6) & (7) & (8) & (9) & (10) & (11) & (12) & (13) & (14) \\
\hline
1 & JPSG031.925+00.083 & 31.925 & 0.083 & 17 & 7 & 185 & 0.37 & 16.53 & 2.50 & 2.56 & 0.12 & 3.88e+02 & 4.35e+02 & 1.12 \\
2 & JPSG031.929+00.056 & 31.929 & 0.056 & 12 & 8 & 151 & 0.34 & 18.23 & 1.71 & 2.42 & 0.08 & 2.66e+02 & 4.35e+02 & 1.64 \\
3 & JPSG031.933+00.090 & 31.933 & 0.090 & 17 & 8 & 244 & 0.40 & 16.38 & 2.62 & 2.19 & 0.12 & 4.06e+02 & 4.94e+02 & 1.22 \\
4 & JPSG031.945+00.076 & 31.945 & 0.076 & 17 & 12 & 211 & 0.49 & 14.99 & 3.92 & 2.97 & 0.18 & 1.01e+03 & 6.45e+02 & 0.64 \\
5 & JPSG031.951+00.071 & 31.951 & 0.071 & 10 & 5 & 148 & 0.24 & 15.31 & 3.60 & 9.01 & 0.17 & 3.72e+02 & 1.66e+02 & 0.45 \\
6 & JPSG031.961+00.066 & 31.961 & 0.066 & 19 & 11 & 97 & 0.50 & 15.58 & 3.49 & 3.07 & 0.16 & 1.08e+03 & 7.45e+02 & 0.69 \\
7 & JPSG031.965+00.062 & 31.965 & 0.062 & 9 & 6 & 116 & 0.25 & 15.60 & 3.53 & 7.87 & 0.17 & 3.64e+02 & 1.95e+02 & 0.54 \\
8 & JPSG031.970+00.060 & 31.970 & 0.060 & 14 & 10 & 255 & 0.41 & 15.67 & 3.43 & 4.59 & 0.16 & 8.87e+02 & 5.08e+02 & 0.57 \\
9 & JPSG031.984+00.065 & 31.984 & 0.065 & 14 & 12 & 231 & 0.44 & 15.36 & 4.32 & 3.51 & 0.20 & 8.93e+02 & 6.76e+02 & 0.76 \\
10 & JPSG031.991+00.067 & 31.991 & 0.067 & 12 & 6 & 260 & 0.29 & 15.15 & 4.78 & 6.93 & 0.22 & 4.94e+02 & 2.94e+02 & 0.60 \\
11 & JPSG031.997+00.067 & 31.997 & 0.067 & 10 & 10 & 110 & 0.34 & 15.12 & 4.94 & 4.37 & 0.23 & 5.11e+02 & 4.17e+02 & 0.82 \\
12 & JPSG032.012+00.057 & 32.012 & 0.057 & 17 & 13 & 205 & 0.51 & 14.00 & 10.85 & 8.76 & 0.51 & 3.36e+03 & 1.23e+03 & 0.37 \\
13 & JPSG032.019+00.065 & 32.019 & 0.065 & 22 & 16 & 144 & 0.64 & 14.92 & 10.02 & 7.39 & 0.47 & 5.69e+03 & 2.64e+03 & 0.46 \\
14 & JPSG032.027+00.059 & 32.027 & 0.059 & 20 & 11 & 95 & 0.51 & 15.39 & 10.89 & 8.86 & 0.51 & 3.37e+03 & 2.13e+03 & 0.63 \\
15 & JPSG032.037+00.056 & 32.037 & 0.056 & 17 & 10 & 262 & 0.45 & 17.59 & 11.49 & 13.76 & 0.54 & 3.56e+03 & 3.75e+03 & 1.05 \\
16 & JPSG032.044+00.059 & 32.044 & 0.059 & 16 & 13 & 109 & 0.49 & 22.32 & 8.08 & 7.15 & 0.38 & 2.50e+03 & 1.33e+04 & 5.30 \\
17 & JPSG032.062+00.071 & 32.062 & 0.071 & 10 & 6 & 132 & 0.27 & 19.62 & 2.03 & 3.86 & 0.09 & 2.09e+02 & 4.96e+02 & 2.37 \\
18 & JPSG032.087+00.076 & 32.087 & 0.076 & 21 & 7 & 237 & 0.42 & 20.53 & 1.55 & 1.54 & 0.07 & 3.20e+02 & 1.22e+03 & 3.83 \\
19 & JPSG032.097+00.076 & 32.097 & 0.076 & 11 & 7 & 154 & 0.30 & 20.66 & 1.58 & 3.10 & 0.07 & 2.44e+02 & 6.78e+02 & 2.78 \\
20 & JPSG032.116+00.090 & 32.116 & 0.090 & 25 & 18 & 150 & 0.73 & 29.94 & 1.74 & 1.05 & 0.08 & 1.17e+03 & 3.92e+04 & 33.49 \\
21 & JPSG032.141+00.172 & 32.141 & 0.172 & 14 & 8 & 240 & 0.36 & 18.85 & 1.67 & 2.50 & 0.08 & 3.46e+02 & 6.06e+02 & 1.75 \\
22 & JPSG032.150+00.133 & 32.150 & 0.133 & 22 & 17 & 262 & 0.66 & 22.69 & 4.62 & 3.11 & 0.22 & 2.62e+03 & 1.59e+04 & 6.07 \\
23 & JPSG032.152+00.122 & 32.152 & 0.122 & 16 & 12 & 125 & 0.47 & 20.09 & 3.42 & 2.84 & 0.16 & 8.83e+02 & 3.01e+03 & 3.41 \\
24 & JPSG032.160+00.113 & 32.160 & 0.113 & 17 & 11 & 209 & 0.47 & 19.73 & 2.00 & 1.73 & 0.09 & 5.15e+02 & 1.57e+03 & 3.05 \\
25 & JPSG032.164+00.171 & 32.164 & 0.171 & 9 & 6 & 216 & 0.25 & 17.86 & 2.12 & 2.37 & 0.10 & 1.10e+02 & 2.67e+02 & 2.43 \\
26 & JPSG032.170+00.135 & 32.170 & 0.135 & 9 & 6 & 163 & 0.25 & 19.23 & 2.19 & 2.44 & 0.10 & 1.13e+02 & 4.27e+02 & 3.77 \\
27 & JPSG032.179+00.137 & 32.179 & 0.137 & 11 & 5 & 147 & 0.25 & 18.54 & 1.97 & 2.13 & 0.09 & 1.02e+02 & 3.16e+02 & 3.10 \\
\hline
\end{tabular}
\end{table}
\end{landscape}

\subsection{Distribution of YSOs}
	\label{sec:YSOs}

\begin{figure*}
	\centering
	\includegraphics[width=0.6\textwidth,trim=8 30 5 60,clip]{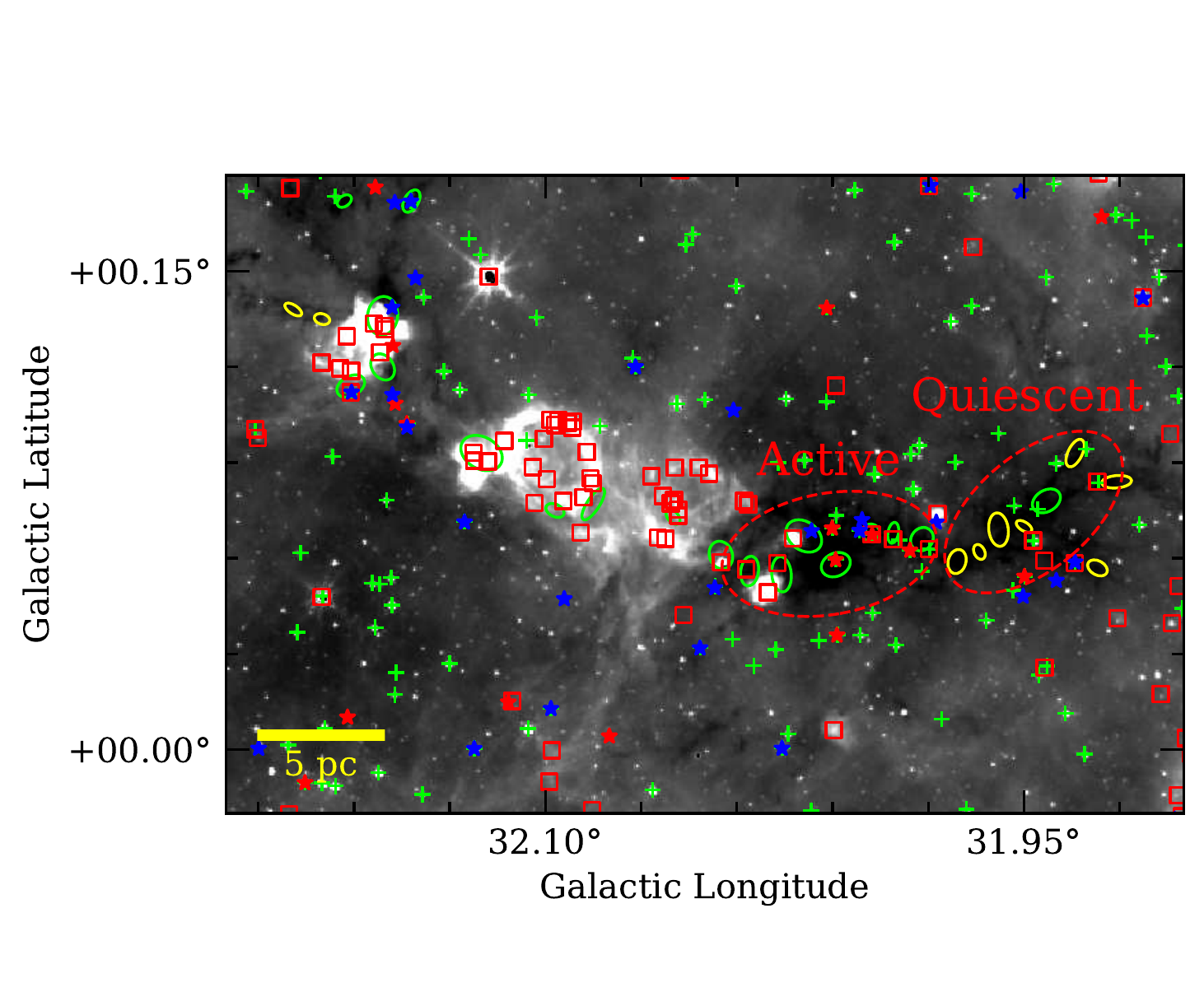}
    \caption{The distributions of Class I YSOs (red stars), Class II YSOs (blue stars), 24\,\micron \ point sources (green crosses) and 70\,\micron \ point sources (red squares). The background image is \emph{Spitzer} 8\,\micron \ continuum map. Starless clumps are shown in yellow ellipses and protostellar clumps are shown in green ellipses. The IRDC can be divided into active part and quiescent part, both of them are marked by red dash ellipses}
    \label{fig:YSOs}
\end{figure*}

We selected sources with 3.6, 4.5, 5.8, 8.0\,\micron \ emission from the GLIMPSE I point-source catalogue \citep{GLIMPSE2009} to identify young stellar objects (YSOs). Classification was performed using the method given in Appendix A of \cite{Gutermuth2009}. 
In brief, the method uses flux ratios or colours, to classify YSO candidates into Class I and Class II, after excluding contamination like star-forming galaxies, broad-line active galactic nuclei (AGNs), and unresolved knots of shock emission. The result of the YSO identification is shown in Figure~\ref{fig:YSOs}. 
Also, the 24\,\micron \ point sources and 70\,\micron \ point sources can be considered as the signature of early-stage star formation, hence, we plot them in Figure~\ref{fig:YSOs} to show the star-formation process in the area more completely. 24\,\micron \ point sources come from \citet{Gutermuth2015} and 70\,\micron \ point sources come from \citet{Molinari2016}.

We classify JPS clumps in the filament into two groups: protostellar clumps, which are associated with any YSOs, 24\,\micron \ or 70\,\micron \ point sources, and starless clumps, which are not. These two groups of clumps are shown in Figure~\ref{fig:YSOs}, with protostellar clumps as green ellipses and starless clumps as yellow ellipses. 
We identify 9 starless clumps and 18 protostellar clumps. Most of starless clumps are located along the IRDC, while most of protostellar clumps are distributed in the active star-forming area.
The distribution of JPS clumps indicates that, along the filament, the evolutionary stages of star formation vary from the east to the west. The \HII regions and IR bubbles which are more highly evolved are located in the east of the filament, while the west of the filament is quiescent.
The IRDC can be divided into two parts, the \emph{active part} and the \emph{quiescent part} as mentioned in \citet{Battersby2014a}. The active part, containing plenty of YSOs and mid-IR point sources, displays signs of active star formation. While the quiescent part, in which most of clumps are starless, shows weak star-forming activity at present.
Most of the protostellar clumps are distributed in the highly evolved part (around \HII regions or in active part of the IRDC). But, in our work, it is hard to distinguish whether the star formation in those clumps is triggered or pre-existing. 

\begin{table*}
	\caption{Physical properties statistics of JPS clumps}
    \label{tab:stat}
\begin{tabular}{*{9}{c}}
\hline
 & $\mathrm{r}_\mathrm{eq}$ & $\mathrm{T}_\mathrm{dust}$ & $\mathrm{N}_{\mathrm{H}_2}$ & $\mathrm{n}_{\mathrm{H}_2}$ & $\Sigma_\mathrm{mass}$ & Mass & Luminosity & L/M ratio \\
 & pc & K & $10^{22}\,\mathrm{cm}^{-2}$ & $10^4\,\mathrm{cm}^{-3}$ & $\mathrm{g}\,\mathrm{cm}^{-2}$ & $\mathrm{M}_{\sun}$ & $\mathrm{L}_{\sun}$ & $\mathrm{L}_{\sun} / \mathrm{M}_{\sun}$ \\
\hline
\multicolumn{9}{c}{Starless clumps} \\
\hline
Min & 0.24 & 10.8 & 1.43 & 1.31 & 0.07 & 6.51e+01 & 3.53e+01 & 0.14 \\
Max & 0.5 & 18.1 & 6.22 & 6.24 & 0.29 & 6.11e+02 & 2.32e+02 & 3.07 \\
Median & 0.34 & 12.72 & 2.86 & 1.73 & 0.13 & 2.57e+02 & 1.03e+02 & 0.37 \\
Mean & 0.33 & 13.48 & 3.12 & 2.42 & 0.15 & 2.67e+02 & 1.17e+02 & 0.80 \\
\hline
\multicolumn{9}{c}{Protostellar clumps} \\
\hline
Min & 0.25 & 11.68 & 0.64 & 0.38 & 0.03 & 4.8e+01 & 7.16e+01 & 0.22 \\
Max & 0.73 & 38.18 & 11.6 & 6.31 & 0.54 & 1.99e+03 & 1.30e+05 & 233.60 \\
Median & 0.46 & 17.05 & 2.875 & 1.66 & 0.14 & 4.66e+02 & 7.88e+02 & 2.00 \\
Mean & 0.45 & 19.64 & 3.98 & 2.14 & 0.19 & 6.71e+02 & 1.42e+04 & 24.47 \\
\hline
\end{tabular}
\end{table*}

The physical properties statistics for both starless clumps and protostellar clumps are presented in Table~\ref{tab:stat}. Starless clumps have lower temperature. And the average luminosity-mass ratio of starless clumps is less than 1, while that of protostellar clumps is about 24.5.

\section{DISCUSSION}
\label{sec:Discussion}
\subsection{Dynamical Structure of IRDC G31.97+0.07}

\subsubsection{Column Density Probability Distribution Functions(PDFs)}
	\label{sec:PDF}

\begin{figure}
	\centering
	\includegraphics[width=\columnwidth,trim=8 20 5 70,clip]{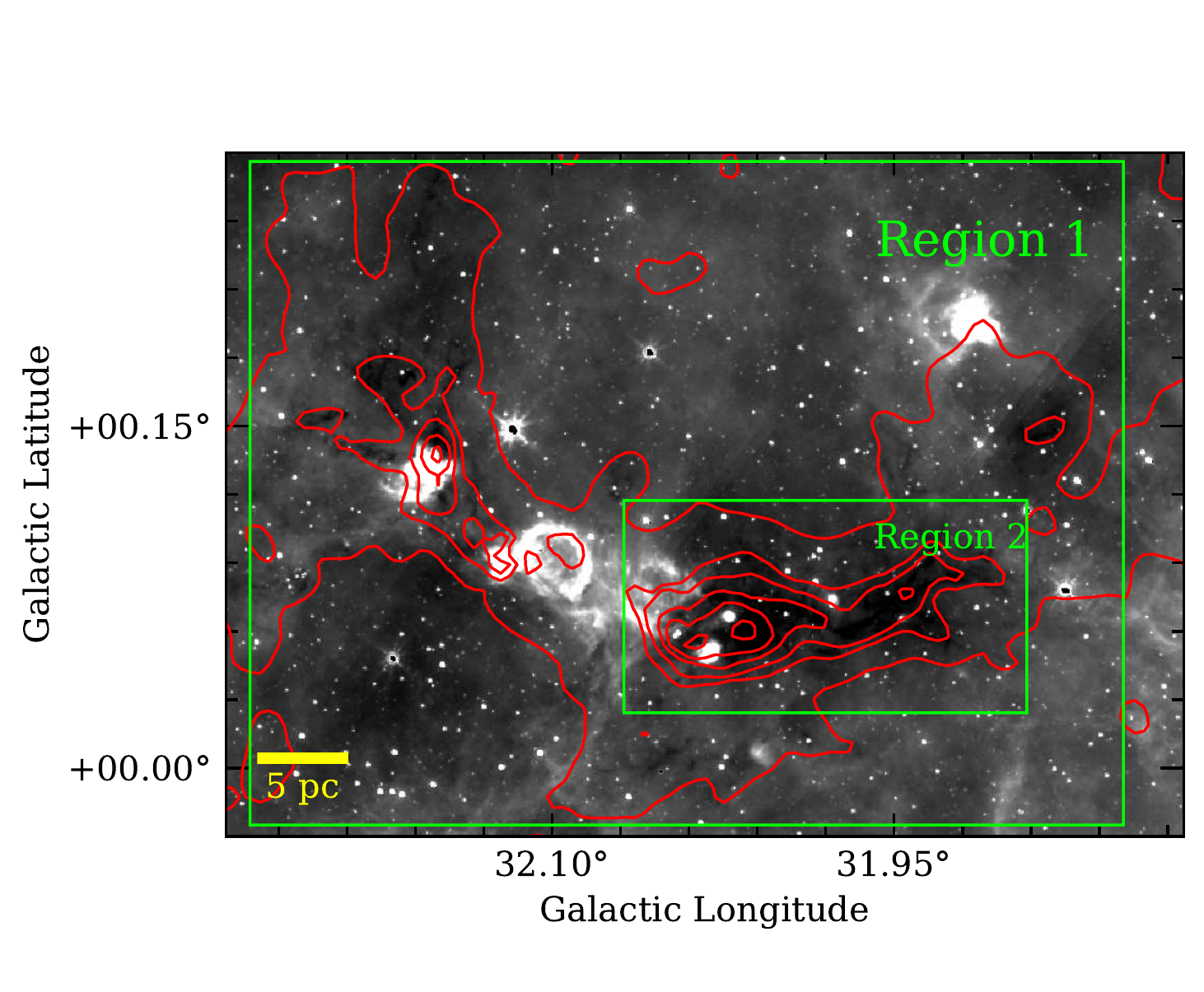}
    \caption{The green rectangles indicates the regions used for PDFs. The background image is \Spitzer \ 8\,\micron \ continuum overlaid by the contour of the column density map derived from pixel-to-pixel SED fitting (see in section~\ref{sec:SED}). Region 1: The whole filamentary structure; Region 2: IRDC G31.97+0.07}
    \label{fig:PDF_region}
\end{figure}

The column density probability distribution function (PDF) is defined as the probability of finding gas within a bin [N,N+dN]. PDFs are a useful tool to study the properties of ISM \citep{Padoan1997ApJ,Burkhart2013}, the core and stellar IMF \citep{Padoan2002,Elmegreen2011,Veltchev2011,Donkov2012}, and the SFR \citep{Krumholz&McKee2005,Padoan&Nordlund2011,Federrath&Klessen2012}. 

Theoretical work and simulations \citep{Vazquez-Semadeni1994,Padoan1997MNRAS,Kritsuk2007,Vazquez-Semadeni2008MNRAS,Federrath2008ApJ,Federrath2010A&A} show that the density PDF of the gas dominated by isothermal supersonic turbulence is approximated well by a log-normal form.
Deviations from the log-normal shape, which are mostly in the form of power-law tails, are present in simulations when compressible turbulence and/or self-gravity are considered \citep{Passot1998,Klessen2000ApJ,Kritsuk2011,Federrath2013ApJ}. 

As many previous studies did, due to the huge range of column densities, we switch to a logarithmic scale $\eta = \ln(N_\mathrm{H_2}) / \langle N_\mathrm{H_2} \rangle$, where $N_\mathrm{H_2}$ is the column density and $\langle N_\mathrm{H_2} \rangle$ is the average column density of the cloud.
So, the form of the PDF at low-densities which is characterized by being log-normal can be written as: 
\begin{equation}
	\label{eq:log-normal}
	p(\eta) = \frac{p_0}{\sqrt{2 \pi \sigma^2_0}} \exp \left( \frac{-(\eta-\eta_0)^2}{2\sigma^2_1} \right),
\end{equation}
where $\eta_0$, $p_0$ and $\sigma_0$ are the peak value, integral probability and standard deviation.
And the power-law tail at high-densities:
\begin{equation}
	\label{eq:power-law}
		p(\eta) \propto N^m_{\rm H_2},
\end{equation}
The models including self-gravity \citep{Klessen2000ApJ,Kritsuk2011,Federrath2013ApJ} indicate that the exponent $m$ can respond to an equivalent spherical density profile of exponent $\alpha$: $\rho(r) \propto r^{-\alpha}$, where $\alpha=1-2/m$. Those work also suggest that  $\alpha$ of a spherical self-gravitating cloud value between 1.5 and 2 ($-4 < m < -2$), which is supported by observational study of several IRDCs \citep{Schneider2015II,Yuan2018ApJ}.

\begin{figure}
	\centering
	\includegraphics[width=0.75\columnwidth,trim=2 0 5 5,clip]{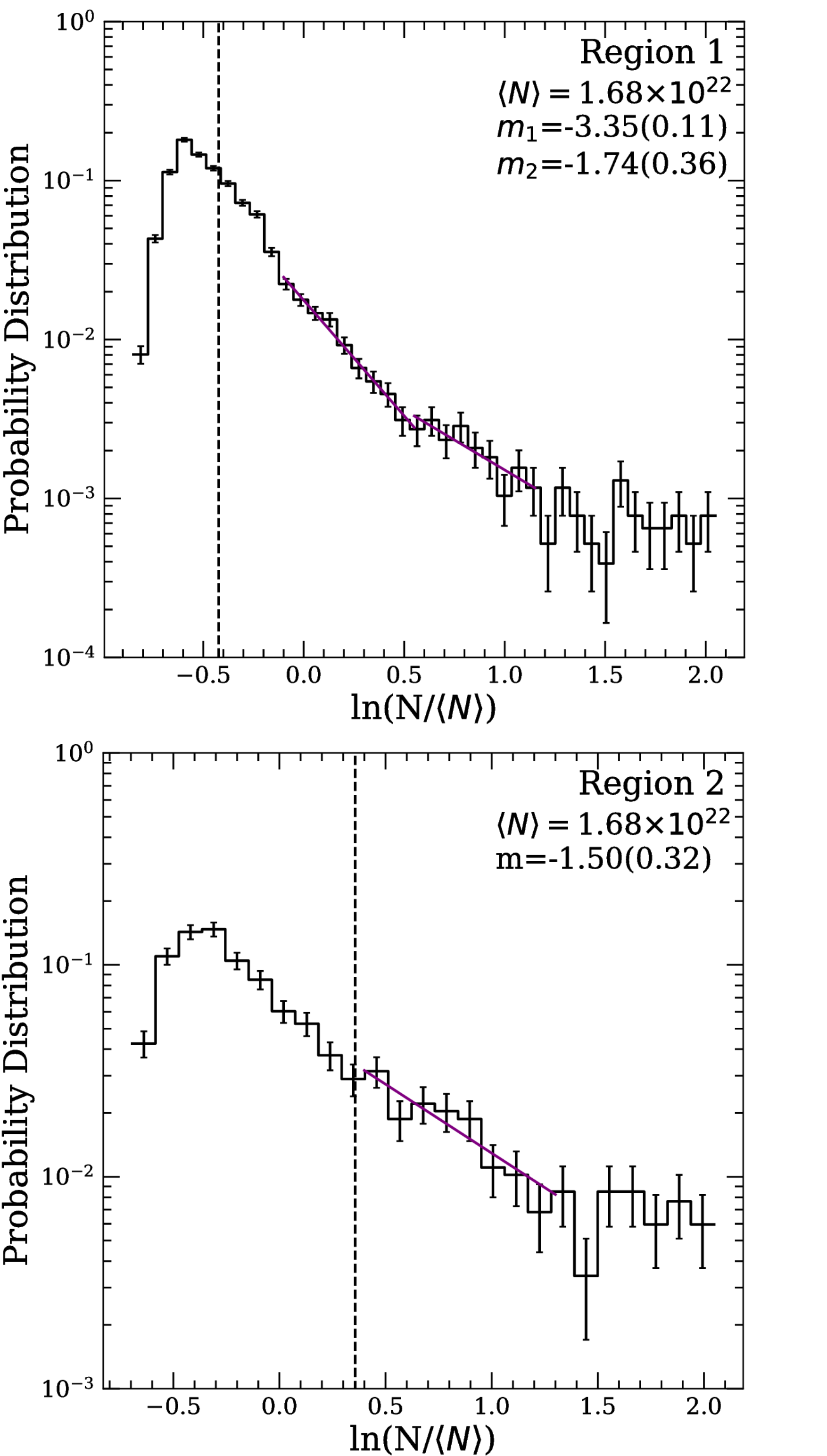}
    \caption{Column density PDFs of the 2 regions shown in Figure~\ref{fig:PDF_region}. The error bars are calculated from Poisson noise in each bin. The black dashed line corresponds to the last closed contour which is regarded as the completeness limit. The power-law tail at high column densities is shown by the violet line.}
    \label{fig:PDF}
\end{figure}

In this work, we determined the PDFs in two regions which are shown in Figure~\ref{fig:PDF_region}. Region 1 covers the whole filamentary structure including the \HII regions and the IRDC. Region 2 only contains the IRDC in the west. 
Figure~\ref{fig:PDF} shows the PDFs of those two regions. The error bars show the statistical Poisson error in each bin. The average column density $\langle N \rangle = 1.68 \times 10^{22}\,\mathrm{cm}^{-2}$.
The black dashed line corresponds to the last closed contour which is regarded as the completeness limit \citep{Lombardi2015,Ossenkopf2016,Alves2017}.
The completeness limit is $-0.42$ for Region 1, and $0.36$ for Region 2.

For the part above the completeness limit, neither PDFs of Region 1 nor Region 2 represent log-normal well. Thus we only discuss their power-law tails.
One can easily notice that the PDF of Region 1 can be described by two power-laws with demarcation between 0.5 and 0.6.
The slope $m$ is $-3.35 (0.11)$ for the power-law at lower column densities and $-1.74 (0.36)$ for the power-law at higher column densities.
We fit the power-law for the PDF of Region 2 from 0.5 to 1.2, the slope $m$ is $-1.50 (0.32)$.
Comparing the PDFs of Region 1 and Region 2, we suggest that flat power-law at higher column densities is mainly contributed from the IRDC.
That means more dense gas accumulate in the IRDC.
The power-law of the PDF of the IRDC is flatter than theoretical prediction, which implies that this region may be compressed by the adjacent \HII region at the south west according Figure~\ref{fig:RGB}.
This suggests that when the ionized-gas pressure is higher than the turbulent ram pressure the ionized-gas would compress the material around the ionizing source, as shown in the numerical simulation of \citet{Tremblin2012A&A}.
And the ionization compression can also play a role at high column densities which are expected to be gravity-dominant, as found by \citet{Tremblin2014A&A} in Rosette nebula.

\subsubsection{Investigating of the fragmentation of the IRDC}
	\label{sec:fragmentation}
The model of a self-gravitating fluid cylinder can be used to describe a filamentary IRDC. The theoretical work describing the fragmentation of fluid cylinders due to the ``sausage" instability has been established decades earlier (e.g., \citealt{Chandrasekhar1953ApJ,Nagasawa1987,Inutsuka1992,Tomisaka1995}).
The theory predicts that the filament should fragment into multiple cores with quasi-regular spacing, that has been observed in several filamentary molecular clouds (e.g., \citealt{Jackson2010ApJ,Wang2014MNRAS,Henshaw2016MNRAS}). The characteristic spacing between cores corresponds to the wavelength of the fastest-growing unstable mode of the fluid instability.

For an incompressible fluid cylinder, the fragment spacing is $\lambda = 11 R$, where $R$ is the radius of the cylinder, given by \citet{Chandrasekhar1953ApJ}. And for an infinite isothermal gas cylinder \citep{Nagasawa1987}:
\begin{equation}
	\label{eq:spacing}
    \lambda = 22 H = \frac{22c_{\rm s}}{(4 \pi G \rho)^{1/2}},
\end{equation}
where $H$ is the isothermal scale height, $c_{\rm s}$ is the sound speed, $G$ is the gravitational constant, and $\rho$ is the gas density of the filament. For a finite isothermal cylinder embedded in external uniform medium, the fragments spacing depend on the ratio of cylinder radius R and isothermal scale height H. If $R \gg H$, $\lambda = 22H$, and if $R \ll H$, $\lambda = 11R$.

For IRDC G31.97+0.07, the sound speed $c_{\rm s} = [(k_{\rm B} T)/(\mu_{\rm p} m_{\rm H})]^{1/2} \approx 0.22\,\mathrm{km\,s}^{-1}$, where $k_{\rm B}$ is the Boltzmann constant, $T\approx 16\,\mathrm{K}$ is the mean temperature derived from SED fitting, $\mu_{\rm p}=2.33$ is the mean molecular weight per free particle, and $m_{\rm H}$ is the mass of atomic hydrogen. 
Thus, the isothermal scale height $H=c_{\rm s} / (4 \pi G \rho)^{1/2} \approx 0.015\,\mathrm{pc}$, where $\rho=\mu_{\rm H_2} m_{\rm H} n_{\rm H_2}$, and $n_{\rm H_2} \sim 5.6 \times 10^4 \, \mathrm{cm}^{-3}$.
Since the radius $R$ of the filament ($\sim 0.01 \degr$, or $\sim 1.2\,\mathrm{pc}$ at $7.07$\,kpc) is much larger than the isothermal scale height $H$, under the isothermal and thermally supported assumption, the predicted theoretical fragment spacing is $\lambda = 22 H = 0.33\,\mathrm{pc}$.

The spacings of 15 clumps located in the IRDC range from $0.66$ to $1.31$\,pc, with an average of $0.97$\,pc and a median of $0.94$\,pc.
This agrees with 
The distribution of the clump spacings is roughly symmetrical and 5 clumps have spacing value in the interval of $[0.89,0.98]$\,pc.
Although the discussion of the clump spacing distribution is difficult for such a small sample,  we can still consider that the clumps have a characteristic spacing of $0.97 \pm 0.03$\,pc and use it in the following discussion.

As we consider the effect of inclination, assuming the inclination angle $i=30\degr$, the observational fragments spacing should be corrected by: $\lambda_{\mathrm{obs},i} = \lambda_\mathrm{obs} / \cos(i) = 1.12$\,pc, which is much larger than the theoretical predicted $0.33$\,pc. 
This discrepancy may be caused by the fact that the IRDC is turbulence and not thermal-pressure dominated.

\begin{figure}
	\centering
    \includegraphics[width=\columnwidth,trim=6 25 6 25,clip]{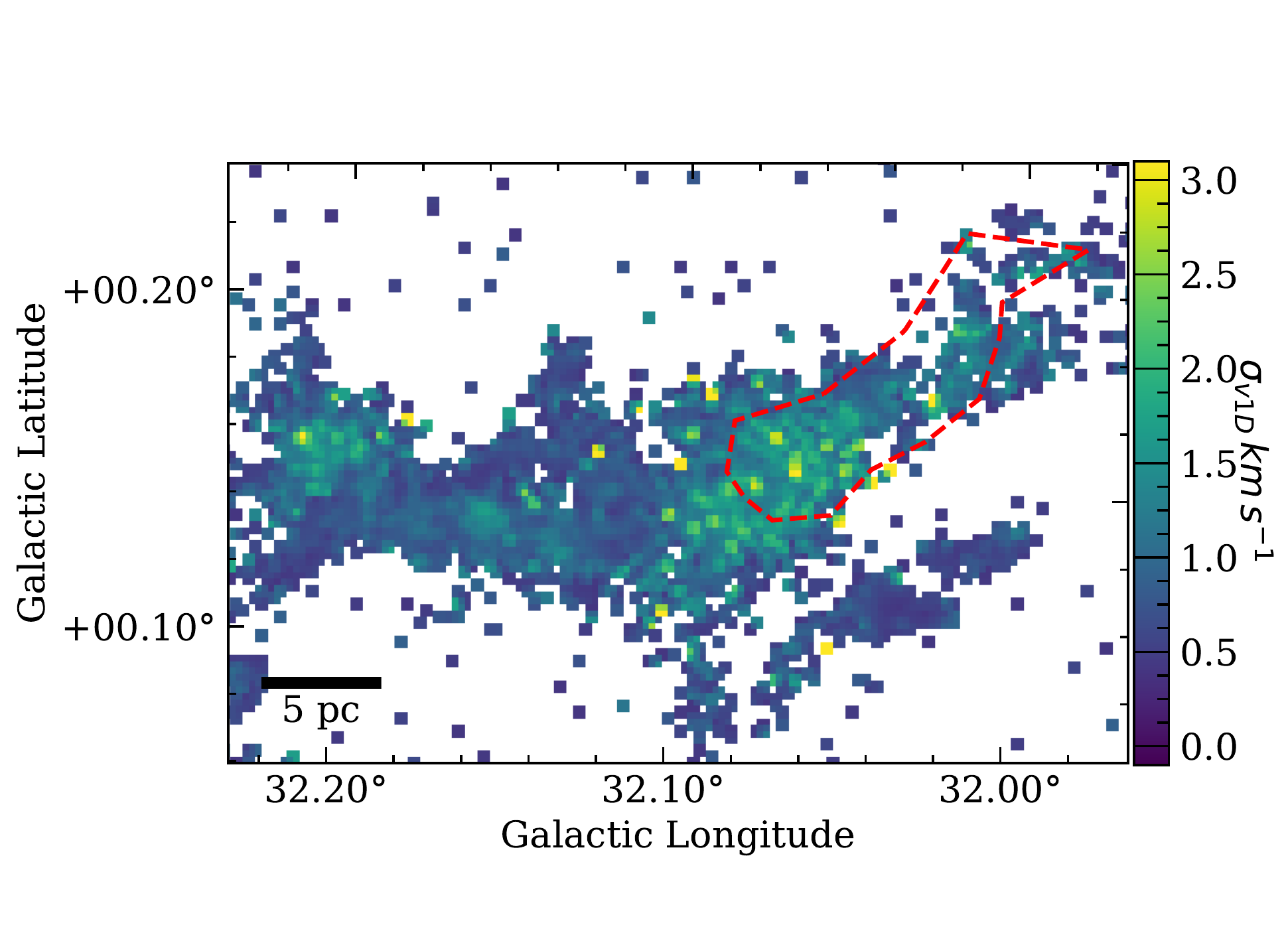}
    \caption{The map of $^{13}$CO (3-2) line velocity dispersion. $\sigma_{v1D}$ of each pixel is derived from Gaussian fitting of the $^{13}$CO (3-2) line observed by JCMT in the velocity interval between 90 and 100\,km\,s$^{-1}$. The red dashed polygon shows the position of the IRDC.}
    \label{fig:vel_dispersion}
\end{figure}

When considering turbulence, the sound speed $c_{\rm s}$ in Equation~\ref{eq:spacing} can be replaced by total velocity dispersion $\sigma_v$ \citep{Fiege2000MNRAS}. To derive the total velocity dispersion of the IRDC, we fitted the spectrum of $^{13}$CO (3-2), observed by JCMT in the velocity interval between 90 and 100\,km\,s$^{-1}$, pixel-by-pixel with a Gaussian profile. The result is shown in Figure~\ref{fig:vel_dispersion}.
The velocity dispersion $\sigma_v$ is given by \citet{Fuller1992ApJ}:
\begin{equation}
	\label{eq:velocity dispersion}
    \sigma_v = \sqrt{ \sigma_{^{13}\mathrm{CO}}^2 + k_B T \left( \frac{1}{\mu_{\rm H_2}m_{\rm H}} - \frac{1}{m_{^{13}\mathrm{CO}}} \right) },
\end{equation}
where $\sigma_{^{13}\mathrm{CO}}$ is the $^{13}$CO line velocity dispersion derived from Gaussian fitting, $m_{^{13}\mathrm{CO}} = 29m_{\rm H}$ is the mass of $^{13}$CO.
For the IRDC, the average $^{13}$CO line velocity dispersion $\sigma_{^{13}\mathrm{CO}} \approx 1.09 \pm 0.45 \,\mathrm{km\,s}^{-1}$, hence, the total velocity dispersion $\sigma_v \approx 1.10 \pm 0.45 \,\mathrm{km\,s}^{-1}$.
Replacing $c_{\rm s}$ in Equation~\ref{eq:spacing} by $\sigma_v$, the expected fragment spacing is $\lambda \approx 1.7 \pm 0.7 \, \mathrm{pc}$, about 1.5 times larger than the observed value but consistent with the error range.

\citet{Kainulainen2013A&A} suggests that at the scale smaller than Jeans' length, the fragmentation more closely resembles Jeans' fragmentation. According to the spherical Jeans' instability, the clump separation is related to density via Jeans' length, $\lambda_\mathrm{J} = c_\mathrm{s} (\pi/(G\rho)) ^{1/2}$. Here the sound speed $c_\mathrm{s}$ is replaced by velocity dispersion $\sigma_v$, and we obtain $\lambda_\mathrm{J} = 0.48$\,pc for the IRDC. That is smaller than all the observational clump spacings, which suggests that the IRDC fragmentation is roughly dominated by the collapse of the natal filament.

The previous discussion is based on the hydrostatic models, but actually, the filament may be non-equilibrium and fragmenting with accreting. 
\citet{Clarke2016} studied the fragmentation of non-equilibrium, accreting  filaments and suggested that an age limit for a filament, which is fragmenting periodically, could be estimated by measuring the average core separation distance $\lambda$ (see also \citealp{Williams2018}):
\begin{equation}
	\tau_\mathrm{age} \geq \tau_\mathrm{crit} \simeq \frac{\lambda}{2c_s},
\end{equation} 
where sound speed $c_s$ should be replaced by velocity dispersion $\sigma_v$. And $\tau_\mathrm{crit}$ can also be related to the mass accretion rate $\dot{M}$:
\begin{equation}
	\tau_\mathrm{crit} = \frac{\lambda}{2 \sigma_v} = \frac{M_\mathrm{line,crit}}{\dot{M}},
\end{equation}
where $M_\mathrm{line,crit} = 2 \sigma_v^2 /G = 563\,M_{\sun}\,\mathrm{pc}^{-1}$ in this work ($G$ is the gravitational constant).
Thus we can derive $\dot{M}$ as $1.3 \times 10^3\,M_{\sun}\,\mathrm{pc}^{-1}\,\mathrm{Myr}^{-1}$ (with $i=0 \degr$) or $1.1 \times 10^3\,M_{\sun}\,\mathrm{pc}^{-1}\,\mathrm{Myr}^{-1}$ (with $i=30 \degr$).
If assuming mass accretion rate remained constant during the filament evolution, we derive the age $\tau_\mathrm{age}$ through:
\begin{equation}
	\tau_\mathrm{age} = \frac{M_\mathrm{line}}{\dot{M}}.
\end{equation}
As the mass of the $11$\,pc long filament is about $7.8\times10^4\,M_{\sun}$ (mentioned at Sec.~\ref{sec:SED}).
The age of filament $\tau_\mathrm{age}$ is $\sim 5.5$\,Myr (with $i=0 \degr$) or $\sim 6.4$\,Myr (with $i=30 \degr$).

%

\subsection{Properties of JPS Clumps}
\subsubsection{CO Line Profiles of Clumps}
	\label{sec:profile}
For each clumps, we extract $^{12}$CO (3-2) spectra from COHRS, $^{13}$CO (3-2) and C$^{18}$O (3-2) from CHIMPS.
The spectra of all clumps are shown in Appendix~\ref{sec:appendix_A}.
Some clumps contain multiple velocity components, which makes our analysis difficult.
The peak values of $^{13}$CO (3-2) for some other clumps do not achieve $3\sigma$ threshold of main beam efficiency corrected CHIMPS data ($\sim 2.4\,$K).
Thus, in the following discussion, we only consider the strongest component of which the $^{13}$CO (3-2) intensity must be higher than $3\sigma$ threshold for each clump, as some previous study did \citep{Urquhart2007,Eden2012,Eden2013}. There are 18 clumps that meet this requirement, 16 of which are protostellar and 2 are starless. And C$^{18}$O (3-2) emission higher than $2.4\,$K are detected in only 9 clumps.

According to radiative transfer, brightness temperature ($T_r$) can be expressed as:
\begin{equation}
   	\label{eq:T_b}
	T_r = \frac{h\nu}{k} [\frac{1}{\exp(h\nu/kT_\mathrm{ex})}-\frac{1}{\exp(h\nu/kT_\mathrm{bg})}] \times (1-e^{-\tau})f,
\end{equation}
where $T_\mathrm{bg}=2.73$\,K is the temperature of cosmic background radiation, and $f$ which is the beam-filling factor can be consider to be $1$ here.
Under the LTE condition, the optical depths of CO and $^{13}$CO can be obtain by:
\begin{equation}
	\label{eq:tau}
	\frac{T_{r,\mathrm{CO}}}{T_{r,^{13}\mathrm{CO}}} \approx \frac{1-\exp(-\tau_\mathrm{CO})}{1-\exp(-\tau_{^{13}\mathrm{CO}})},
\end{equation}
here, $\tau_\mathrm{CO}/\tau_{^{13}\mathrm{CO}} = [\mathrm{CO}]/[^{13}\mathrm{CO}] \approx 89$.
According the equation we obtain the optical depths of $^{13}$CO, the results are listed in the Column 7 in the Table~\ref{tab:line_profile}.
The $\tau_{^{13}\mathrm{CO}}$ ranges from $0.33$ to $1.05$, so we can generally consider $^{13}$CO (3-2) as optically thin.

\begin{table*}
\centering
\caption{Line profiles parameters of JPS clumps. Optically thick line: CO (3-2), Optically thin line: $^{13}$CO (3-2). `*' labels the clumps of which C$^{18}$O (3-2) are used as optically thin lines in the following discussion; (2): Evolutionary type of clumps (see in Section~\ref{sec:YSOs}); (3)-(9): parameters used in Equation~\ref{eq:line_profile} to select blue/red profile; (7): line profiles, BP: blue profile, RP: red profile. `**' labels the clumps which may be misclassified by multiple velocity components; (11): Infall velocities of the clumps with blue profiles, obtained from Equation~\ref{eq:infall}.}
\label{tab:line_profile}
\begin{tabular}{cccccccccccc}
\hline
ID & Name & Type & $V_\mathrm{peak,thick}$ & $T_{r,\mathrm{thick}}$ & $V_\mathrm{peak,thin}$ & $T_{r,\mathrm{thin}}$ & $\tau_{^{13}\mathrm{CO}}$ & FWHM$_\mathrm{thin}$ & $\delta_V$ & Profile & $V_\mathrm{in}$\\
 & & & km\,s$^{-1}$ & K & km\,s$^{-1}$ & K & & km\,s$^{-1}$ &  & km\,s$^{-1}$ \\
 & (1) & (2) & (3) & (4) & (5) & (6) & (7) & (8) & (9) & (10) & (11)\\
\hline
4 & JPSG031.945+00.076 & Protostellar & 93.1 & 8.29 & 96.00 & 2.45 & 0.35 &2.96 & -0.98 & BP & 0.21 \\
9 & JPSG031.984+00.065 & Protostellar & 92.1 & 11.85 & 93.68 & 3.88 & 0.40 & 2.52 & -0.63 & BP** & \\
10 & JPSG031.991+00.067 & Protostellar & 92.1 & 14.43 & 94.62 & 4.01 & 0.33 & 4.09 & -0.62 & BP & 0.47\\
11 & JPSG031.997+00.067 & Protostellar & 92.1 & 13.51 & 95.10 & 4.46 & 0.40 & 3.68 & -0.82 & BP & 0.21\\
12 & JPSG032.012+00.057 & Protostellar & 94.1 & 13.75 & 96.31 & 5.03 & 0.45 & 3.31 & -0.67 & BP** & \\
13 & JPSG032.019+00.065 & Protostellar & 97.1 & 14.59 & 98.87 & 5.99 & 0.53 & 4.15 & -0.43 & BP** & \\
14 & JPSG032.027+00.059 & Protostellar & 93.1 & 19.47 & 94.73 & 6.34 & 0.39 & 3.80 & -0.43 & BP** & \\
15 & JPSG032.037+00.056 & Protostellar & 93.1 & 17.54 & 94.49 & 8.08 & 0.62 & 3.26 & -0.43 & BP** & \\
16 & JPSG032.044+00.059 & Protostellar & 92.1 & 16.45 & 94.82 & 10.36 & 0.99 & 3.45 & -0.79 & BP & 0.13\\
17 & JPSG032.062+00.071 & Protostellar & 95.1 & 13.42 & 95.42 & 6.32 & 0.63 & 2.44 & -0.13 & ... & \\
18 & JPSG032.087+00.076 & Protostellar & 97.1 & 14.25 & 97.01 & 8.57 & 0.92 & 2.00 & 0.05 & ... & \\
19 & JPSG032.097+00.076 & Protostellar & 94.1 & 10.80 & 95.98 & 7.04 & 1.05 & 2.58 & -0.73 & BP & 0.44\\
20 & JPSG032.116+00.090* & Protostellar & 97.1 & 22.46 & 96.18 & 6.40 & 0.96 & 2.65 & 0.35 & RP & \\
22 & JPSG032.150+00.133* & Protostellar & 96.1 & 19.77 & 94.45 & 3.35 & 0.72 & 3.94 & 0.42 & RP & \\
23 & JPSG032.152+00.122 & Protostellar & 96.1 & 15.60 & 96.08 & 8.56 & 0.80 & 2.32 & 0.01 & ... & \\
24 & JPSG032.160+00.113 & Protostellar & 96.1 & 14.88 & 96.46 & 7.46 & 0.70 & 1.82 & -0.20 & ... & \\
26 & JPSG032.170+00.135 & Starless & 95.1 & 8.72 & 94.83 & 3.61 & 0.53 & 1.94 & 0.14 & ... & \\
27 & JPSG032.179+00.137 & Starless & 95.1 & 5.92 & 94.94 & 2.70 & 0.61 & 1.67 & 0.09 & ... & \\
\hline
\end{tabular}
\end{table*}

The asymmetry of line profiles can be used to probe different dynamic features. Generally, blue profiles are caused by collapse or infall motion, and red profiles are linked to expansion or outflow motion \citep{Mardones1997ApJ}. 
To select clumps with blue or red profiles, we adopted the normalized criterion set by \citet{Mardones1997ApJ}:
\begin{equation}
	\label{eq:line_profile}
	\delta_V = (V_\mathrm{thick} - V_\mathrm{thin}) / \Delta V_\mathrm{thin},
\end{equation}
where $V_\mathrm{thick}$ is the velocity of the optically thick line (i.e., CO (3-2) in our discussion) at peak value, $V_\mathrm{thin}$ and $\Delta V_\mathrm{thin}$ are the systemic velocity and line width of the optically thin line, i.e., $^{13}$CO (3-2). For the clumps which may contain multiple components, their $\Delta V_\mathrm{thin}$ are measured by Gaussian fitting for the strongest component.
If $\delta_V < -0.25$, the line profile is considered a blue profile, and if $\delta_V > 0.25$, the line profile is considered to be red. 
For Clump 20 and Clump 22, comparing profiles of $^{13}$CO and C$^{18}$O (3-2), we find that their $^{13}$CO (3-2) might be self-absorbed. As their C$^{18}$O (3-2) emissions are strong enough, we choose their C$^{18}$O (3-2) as optically thin lines in this part of discussion.

Our classification is shown in Table~\ref{tab:line_profile}. For the 18 clumps, we find that 10 clumps show blue profiles, and 2 clumps show red profiles. After visual inspection we exclude five clumps as they may be misclassfied by multiple velocity components, and they are marked in Table~\ref{tab:line_profile}.

According to \citet{Myers1996}, we can estimate infall velocities of the clumps with blue profiles:
\begin{equation}
	\label{eq:infall}
	V_\mathrm{in} \approx \frac{\sigma_v}{v_\mathrm{red}-v_\mathrm{blue}} \ln \left( \frac{1+eT_\mathrm{BD}/T_\mathrm{D}}{1+eT_\mathrm{RD}/T_\mathrm{D}} \right),  
\end{equation}
where $T_\mathrm{D}$ is the brightness temperature of the dip of the optically thick line. $T_\mathrm{BD}$($T_\mathrm{RD}$) is the height of the blue (red) peak above the peak. The velocity dispersion $\sigma_v$ is obtained from the FWHM of a optically thin line.

The calculation results of five clumps with blue profiles are listed in Table~\ref{tab:line_profile}. 
The infall velocities of them range from 0.13 to 0.47\,km\,s$^{-1}$, with a mean value of 0.31\,km\,s$^{-1}$.

\subsubsection{The Stability of Clumps}
\label{sec:stability}
Using the physical properties we obtain in the previous analysis, we can study the stability of JPS clumps, i.e., whether they are susceptible to gravitational collapse or expansion in the absence of pressure confinement.

First, we consider the thermal Jeans instability,
\begin{equation}
	\label{eq:Jeans mass}
	M_\mathrm{J} = \frac{\pi^{5/2} c_\mathrm{s}^3}{6 \sqrt{G^3 \rho}}\ , 
\end{equation}
\begin{equation}
	\label{eq:Jeans length}
	\mathrm{and} \ \lambda_\mathrm{J} = c_\mathrm{s} \left( \frac{\pi}{G\rho} \right) ^{1/2},
\end{equation}
where $M_\mathrm{J}$ and $\lambda_\mathrm{J}$ are the Jeans mass and Jeans length, $c_\mathrm{s}$ is the sound speed and $\rho = \mu_{\mathrm{H_2}}m_{\mathrm{H}}n_{\mathrm{H_2}}$ is the density. When clumps are both thermally and turbulently supported, the sound speed $c_\mathrm{s}$ is replaced by the velocity dispersion $\sigma_v$.

Another important parameter to estimate the stability of clumps is virial parameter $\alpha_\mathrm{vir}$:
\begin{equation}
	\label{eq:virial parameter}
	\alpha_\mathrm{vir} \equiv \frac{M_\mathrm{vir}}{M_\mathrm{clump}} = \frac{5 \sigma_v^2 R_\mathrm{eq}}{G M_\mathrm{clump}},
\end{equation}
similarly $\sigma_v$ is the velocity dispersion contributed from both thermal motions and turbulence, $R_\mathrm{eq}$ is the equivalent radius of the clump and $M_\mathrm{clump}$ is the clump mass. According to \citet{McKee1999ApJ} and \cite{Kauffmann2013ApJ}, in a non-magnetized cloud, if $\alpha_\mathrm{vir}<2$, the cloud is gravitationally bound and will tend to collapse, otherwise, it may expand by lack of pressure confinement.

The calculation results are shown in Table~\ref{tab:gas_para}. In our calculation, we adopt the velocity dispersion derived from the width of optically thin line, which have been obtained in Sec.~\ref{sec:profile}.
Only one clump has $\alpha_\mathrm{vir}>2$, suggesting that most the clumps of the filament are gravitationally bound and likely to collapse. There are 12 clumps with $\lambda_\mathrm{Jeans} < 2 R_\mathrm{eq}$, indicating that they are likely to split into smaller fragments while collapsing.
In addition, we have to point out that the $\alpha_\mathrm{vir}$ of three clumps (Clump 10, 11 and 19) with blue profiles may be overestimate, as the line may be broadened by gas infall motion.

\begin{table*}
\centering
\caption{
The gas dynamical properties and viral parameters of JPS clumps derived from $^{13}$CO (3-2) data. (3): The velocity dispersion which is derived from the optically thin line width in Table~\ref{tab:line_profile}. (5)-(8): Viral mass, Jeans mass, Jeans length and viral parameter calculated from velocity dispersion.
}
\label{tab:gas_para}
\begin{tabular}{ccccccccc}
\hline
ID & Name & R$_\mathrm{eq}$ & $\sigma_v$ & M$_\mathrm{clump}$ & M$_\mathrm{vir}$ & M$_\mathrm{Jeans}$ & $\lambda_\mathrm{Jeans}$ & $\alpha_\mathrm{vir}$ \\
 & & pc & km\,s$^{-1}$ & M$_{\sun}$ & M$_{\sun}$ & M$_{\sun}$ & pc & \\
 & (1) & (2) & (3) & (4) & (5) & (6) & (7) & (8) \\
\hline
4 & JPSG031.945+00.076 & 0.49 & 1.27 & 1.01e+03 & 9.24e+02 & 4.71e+02 & 0.76 & 0.92 \\
9 & JPSG031.984+00.065 & 0.44 & 1.09 & 8.93e+02 & 6.07e+02 & 2.71e+02 & 0.60 & 0.68 \\
10 & JPSG031.991+00.067 & 0.29 & 1.75 & 4.94e+02 & 1.03e+03 & 7.95e+02 & 0.68 & 2.08 \\
11 & JPSG031.997+00.067 & 0.34 & 1.57 & 5.11e+02 & 9.79e+02 & 7.32e+02 & 0.77 & 1.92 \\
12 & JPSG032.012+00.057 & 0.51 & 1.42 & 3.36e+03 & 1.19e+03 & 3.80e+02 & 0.49 & 0.36 \\
13 & JPSG032.019+00.065 & 0.64 & 1.77 & 5.69e+03 & 2.34e+03 & 8.08e+02 & 0.67 & 0.41 \\
14 & JPSG032.027+00.059 & 0.51 & 1.63 & 3.37e+03 & 1.57e+03 & 5.69e+02 & 0.56 & 0.47 \\
15 & JPSG032.037+00.056 & 0.45 & 1.40 & 3.56e+03 & 1.03e+03 & 2.92e+02 & 0.39 & 0.29 \\
16 & JPSG032.044+00.059 & 0.49 & 1.49 & 2.50e+03 & 1.26e+03 & 4.82e+02 & 0.57 & 0.50 \\
17 & JPSG032.062+00.071 & 0.27 & 1.06 & 2.09e+02 & 3.52e+02 & 2.38e+02 & 0.55 & 1.68 \\
18 & JPSG032.087+00.076 & 0.42 & 0.88 & 3.20e+02 & 3.80e+02 & 2.18e+02 & 0.73 & 1.19 \\
19 & JPSG032.097+00.076 & 0.30 & 1.12 & 2.44e+02 & 4.36e+02 & 3.12e+02 & 0.65 & 1.79 \\
20 & JPSG032.116+00.090 & 0.73 & 1.38 & 1.17e+03 & 1.61e+03 & 1.00e+03 & 1.38 & 1.38 \\
22 & JPSG032.150+00.133 & 0.66 & 1.77 & 2.62e+03 & 2.40e+03 & 1.23e+03 & 1.03 & 0.91 \\
23 & JPSG032.152+00.122 & 0.47 & 1.01 & 8.83e+02 & 5.59e+02 & 2.41e+02 & 0.62 & 0.63 \\
24 & JPSG032.160+00.113 & 0.47 & 0.81 & 5.15e+02 & 3.57e+02 & 1.58e+02 & 0.63 & 0.69 \\
26 & JPSG032.170+00.135 & 0.25 & 0.85 & 1.13e+02 & 2.12e+02 & 1.56e+02 & 0.56 & 1.87 \\
27 & JPSG032.179+00.137 & 0.25 & 0.74 & 1.02e+02 & 1.61e+02 & 1.11e+02 & 0.52 & 1.58 \\
\hline
\end{tabular}
\end{table*}

\subsubsection{The Evolutionary Stage of Clumps}
The luminosity-mass diagram is an effective tool to infer the evolutionary phases of the clumps. \citet{Saraceno1996A&A} used this diagram to describe the evolutionary stages of low-mass objects, and \citet{Molinari2008A&A} used it extendedly for the study of evolutionary tracks of high-mass regions.

\begin{figure}
	\centering
	\includegraphics[width=0.9\columnwidth,trim=0 0 40 30,clip]{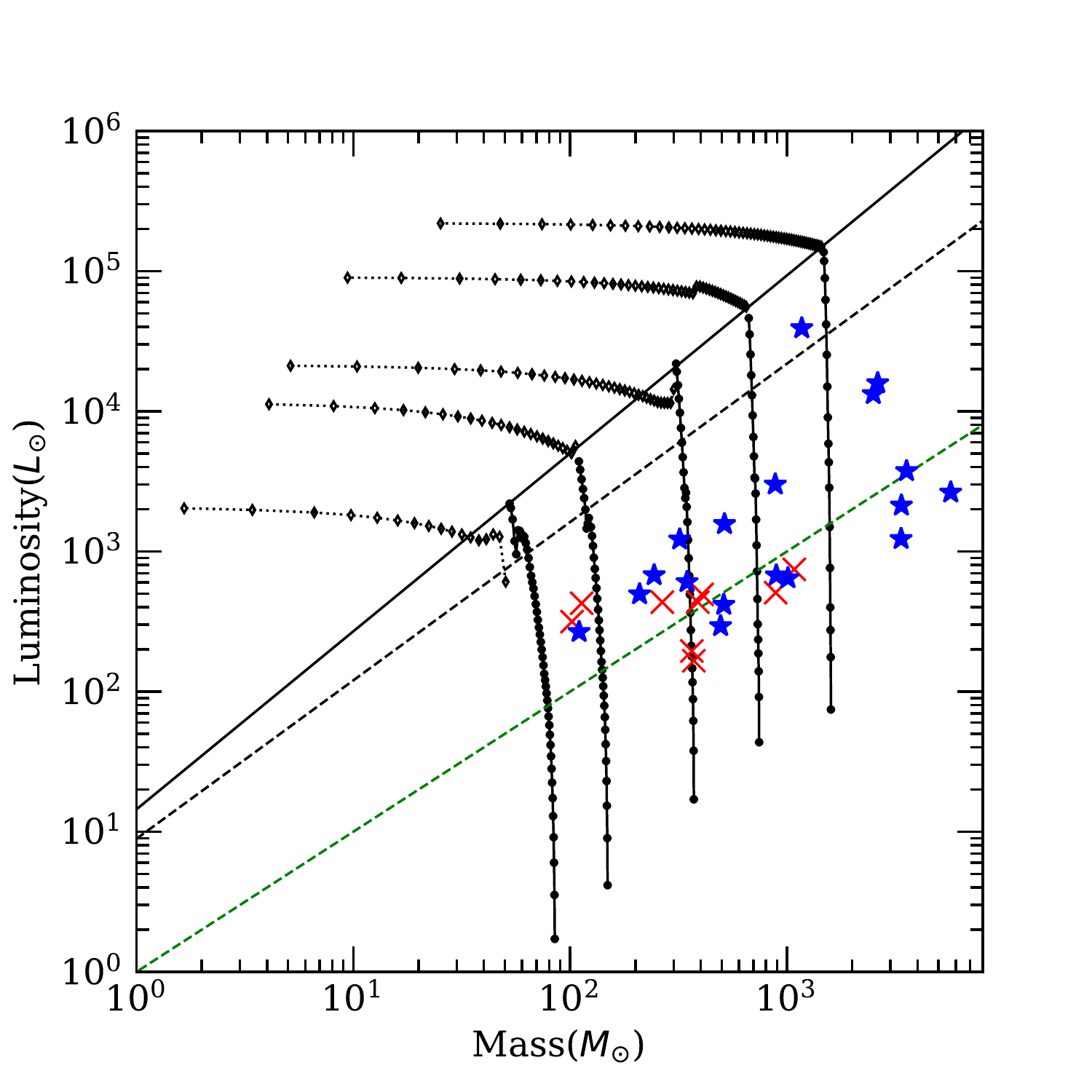}
	\caption{Luminosity-mass diagram for the clumps superimposed with the model of \citet[black tracks with symbols]{Molinari2008A&A}. Starless clumps are shown by red crosses, and protostellar clumps by blue stars. The different tracks (left to right) are for different initial clump masses of 80, 140, 350, 700 and 2000 $M_{\sun}$. The black solid and dashed lines are the best log-log fit for Class I and Class 0 sources extrapolated in the high-mass regime by \citet{Molinari2008A&A}. The green dashed line shows the characteristic of $L_\mathrm{clump} / M_\mathrm{clump} \leq 1\,L_{\sun}/M_{\sun}$ suggested by \citet{Molinari2016ApJ} for starless clumps.
	}
	\label{fig:l-m}
\end{figure}

The luminosity-mass diagram of the clumps is shown in Figure~\ref{fig:l-m} with starless clumps denoted by red crosses and protostellar clumps by blue stars. The diagram is also superimposed with the evolutionary paths from \citet{Molinari2008A&A} for cores with different initial envelope mass.
The paths follow the two-phase model of \citet{McKee2003ApJ}. According to the model, in the first phase, when a clump gravitationally collapses, the mass slightly decreases due to accretion and molecular outflow, while its luminosity significantly increases, and the object moves along an almost vertical path in the $L_\mathrm{clump}-M_\mathrm{clump}$ diagram. At the end of the first phase, the object is surrounded by an \HII region and begins to expel surrounding material through radiation and outflow. In the second phase, the object follows a horizontal track with a nearly constant luminosity and decreasing mass. 
Although these evolutionary tracks are initially modeled for single cores and some high-mass clumps may contain multiple cores in different evolutionary stages, the $L_\mathrm{clump}-M_\mathrm{clump}$ diagram has been used in the past to discuss the evolutionary stages of both low-mass and high-mass clumps \citep{Hennemann2010A&A,Elia2010A&A,Traficante2015MNRAS,Yuan2017ApJS}. 

The black solid and dashed lines in Figure~\ref{fig:l-m} show the best log-log fit for Class I and Class 0 objects. We only find two protostellar clumps located between these two lines, and one protostellar clump lies above the log-log fit of Class I objects. \citet{Molinari2016ApJ} suggest that $L_\mathrm{clump} / M_\mathrm{clump} \leq 1\,L_{\sun} / M_{\sun}$ is the characteristic for starless clumps, and we show this criterion by the green dashed line in Figure~\ref{fig:l-m}. However, \citet{Traficante2015MNRAS} identify 667 starless clumps in IRDCs from \Herschel \ data, and suggest  the mean $L_\mathrm{clump} / M_\mathrm{clump} \simeq 1.1$ for the starless clump distribution. Generally, the starless clumps in our sample are distributed along or under the green dashed line, with mean $L_\mathrm{clump} / M_\mathrm{clump} \simeq 1.4$, which indicates that these clumps are still in the very early stages of their evolution. 
We also find seven protostellar clumps with $L_\mathrm{clump} / M_\mathrm{clump} < 1$. We suggest that they may be misclassified due to  foreground mid-IR point sources. Another probable explanation is that they do contain YSOs, but the rest of the clump is still cold and quiescent. Thus on average, their $L-M$ ratios still keep low.

\subsubsection{High-mass Star Forming Regions}
\begin{figure}
	\centering
	\includegraphics[width=0.9\columnwidth,trim=0 0 40 30,clip]{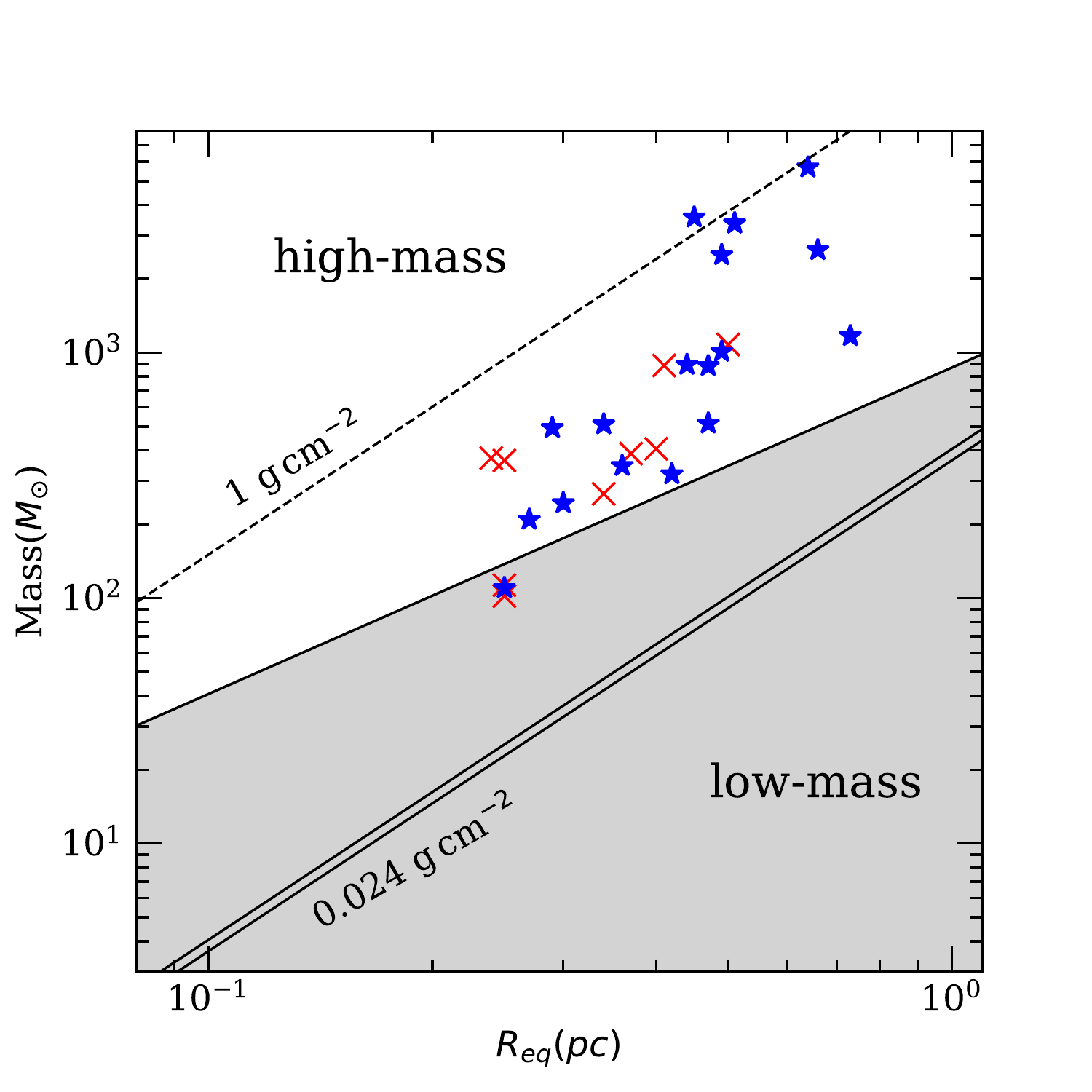}
	\caption{Clump mass as a function of equivalent radius for protostellar clumps (blue stars) and starless clumps (red crosses). The unshaded area delimits the region of high-mass star formation. The threshold is $M \geq 870 \, M_{\sun} (r / \mathrm{pc})^{1.33}$, adopted from \citet{Kauffmann2010ApJb}. Lower solid lines represent surface density thresholds for ``efficient" star formation of $116\,M_{\sun}\,\mathrm{pc}^{-2}$ ($\sim 0.024\,\mathrm{g\,cm}^{-2}$) \citep{Lada2010ApJ} and $129\,M_{\sun}\,\mathrm{pc}^{-2}$ ($\sim 0.027\,\mathrm{g\,cm}^{-2}$) \citep{Heiderman2010ApJ}. The upper dashed line shows the criterion of $1\,\mathrm{g\,cm}^{-2}$ for high-mass star formation given by \citet{Krumholz2008Nature}.
	 }
	 \label{fig:m-r}
\end{figure}

To assess the potential of clumps to form massive stars, we have to consider their sizes and masses. Figure~\ref{fig:m-r} shows the mass versus equivalent radius diagram on which starless clumps are shown by red crosses and protostellar clumps by blue stars. All the JPS clumps satisfy the threshold for ``efficient" star formation of $116\,M_{\sun}\,\mathrm{pc}^{-2}$ ($\sim 0.024\,\mathrm{g\,cm}^{-2}$) given by \citet{Lada2010ApJ} or $129\,M_{\sun}\,\mathrm{pc}^{-2}$ ($\sim 0.027\,\mathrm{g\,cm}^{-2}$) given by \citet{Heiderman2010ApJ}. The thresholds are shown as the lower solid lines in Figure~\ref{fig:m-r}. That means all of the clumps including starless clumps have sufficient material to form stars.

According to observations of nearby clouds, \citet{Kauffmann2010ApJb} give a more restrictive massive-star formation criterion of $M \geq 870 \, M_{\sun} (r / \mathrm{pc})^{1.33}$. Only $4$ clumps don't meet the criterion of massive-star formation, thus most clumps can potentially form high-mass stars. Six of the clumps above the threshold are starless clumps, so they are good candidates for studying massive star-forming regions in a very early evolutionary stage.

Moreover, mass surface density $\Sigma_\mathrm{mass}$ is another useful parameter to identify whether clumps have sufficient mass to form massive stars. \citet{Krumholz2008Nature} suggest that high-mass star formation requires a mass surface density larger than $1\,\mathrm{g\,cm}^{-2}$ to prevent fragmentation into low-mass cores through radiative feedback. Only one clump meets this criterion; however, this threshold is relatively uncertain, and it does not consider magnetic fields, which may play a significant role in preventing fragmentation. In addition, massive clumps and cores with $\Sigma$ less than $1\,\mathrm{g\,cm}^{-2}$ are reported in some observations (e.g. \citealt{Butler2012ApJ,Tan2013ApJ}).

\section{Summary}
\label{sec:Summary}
We present multiple infrared, sub-millimetre continuum and CO isotope rotation-line observations toward IRDC G31.97+0.07 on a large scale. From the continuum and spectral data, we get the dust temperature, column density, excitation temperature and velocity dispersion of this region. The main results of this work are summarized as follows:

\begin{itemize}
	
	\item[1.] The dust temperature and molecular-hydrogen column density maps are derived from fitting the SED of the \Herschel \ data pixel by pixel, using a grey-body model. The dust temperature of the IRDC is lower than the active star-formation region that contains \UCHII \ regions and IR bubbles, while the column density of the IRDC is higher. The total mass is about $2.5\times 10^5\,M_{\sun}$ for the whole filamentary structure and is about $7.8 \times 10^4\,M_{\sun}$ for the IRDC, based on the results.
	
	\item[2.] From the column density map derived from SED fitting, we produce column density probability distribution functions (PDFs) toward two regions, Region 1: the whole filamentary structure; Region 2: the IRDC. 
	The PDFs of both Region 1 and Region 2 mainly show power-law tails at the part above the completeness limit suggesting that this region is gravity-dominant. 	

	Compared to other part, the power-law slope of the PDF of Region 2 is flatter, suggesting that more dense gas accumulate in the IRDC. For the PDF of Region 2, the power-law slope $m=-1.50(0.32)$ is flatter than the prediction of a spherical self-gravitating cloud model (-4<m<-2), which implies that this region may be compressed by the adjacent \HII region.

	\item[3.] The theory of self-gravitating, hydrostatic cylinders shows that the filament would fragment into cores with a roughly constant spacing caused by the ``sausage" instability.
In this work, we derive the mean velocity dispersion by fitting the FWHM of the optically thin line, i.e., $^{13}$CO (3-2). Taking the average velocity dispersion $\sigma_{v} \sim 1.1\,\mathrm{km\,s}^{-1}$, number density $n_\mathrm{\rm H_2} \sim 2.2 \times 10^4\,\mathrm{cm}^{-3}$ and inclination angle $i = 30\degr$, we find the mean observed fragment spacing ($\sim 1.12$\,pc) is about 1.5 times smaller than the prediction of the theoretical model ($\sim 1.7$\,pc).
	
Using another non-equilibrium, accreting model \citep{Clarke2016}, we estimate the age of the IRDC, which is about $6.4\,$Myr when $i = 30\degr$.
	
	\item[4.] There are 27 clumps which are identified from JPS 850\,\micron \ continuum located in the filamentary structure. 
	Using the GLIMPSE point-source catalogue and the identification of the YSOs in the previous study, we get distributions of Class I and Class II YSOs. We classify them into two groups: protostellar clumps, which are associated with any YSOs, 24 or 70\,\micron \ point sources; and starless clumps which are not. We identify 9 starless clumps and 18 protostellar clumps.
	We derive physical properties of all the clumps from SED fitting. Compared to protostellar clumps, starless clumps have lower temperature and luminosity-mass ratios.
	
	\item[5.] Spectral analysis are performed for 18 clumps with relatively strong $^{13}$CO (3-2) emission. Using the optically thick line (i.e $^{12}$CO) and the optically thin (i.e $^{13}$CO or C$^{18}$O) spectra simultaneously, we identify 10 clumps with blue profiles asymmetry, while 5 of them may be misclassified by multiple velocity components. Blue profiles indicate infall motion. We obtain the average infall velocity of the clumps with blue profiles, which is about $0.31\,$km\,s$^{-1}$.
	
	We also calculate the virial parameters of all the 18 clumps to investigate their stabilities. Only one clumps has  $\alpha_\mathrm{vir}>2$, suggesting that most clumps are gravitationally bound and tend to collapse.
	
	\item[6.] All of the 27 JPS clumps fulfil the ``efficient" star-formation threshold, thus they have sufficient mass to form stars. 23 clumps ($\sim$ 85\%) are above the threshold for high-mass star formation proposed by \citet{Kauffmann2010ApJb}, and 6 of them are starless clumps.
	According to the luminosity-mass diagram, all the starless clumps in our sample are in very early evolutionary stages. Hence, those 6 starless clumps are good candidates for studying high-mass star-forming regions in very early evolutionary stage.
\end{itemize}


\section*{Acknowledgements}

This work was supported by the China Ministry of Science and Technology under the State R\&D Program (2017YFA0402600), and by the NSFC grant no. U1531246, 11503035, and also supported by the Open Project Program of the Key Laboratory of FAST, NAOC, Chinese Academy of Sciences. J.H. Yuan is partly supported by the Young Researcher Grant of National Astronomical Observatories, Chinese Academy of Sciences.

The James Clerk Maxwell Telescope has historically been operated by the Joint Astronomy Centre on behalf of the Science and Technology Facilities Council of the United Kingdom, the National Research Council of Canada and the Netherlands Organisation for Scientific Research. 
JCMT continuum data were obtained by SCUBA-2. Additional funds for the construction of SCUBA-2 were provided by the Canada Foundation for Innovation. 
This work made use of data from the \Spitzer Space Telescope, which is operated by the Jet Propulsion Laboratory, California Institute of Technology
under a contract with NASA.
This work also used data from Hi-GAL which is one of key projects of \Herschel spacecraft. The \Herschel spacecraft was designed, built, tested, and launched under a contract to ESA managed by the Herschel/Planck Project team by an industrial consortium under the overall responsibility of the prime contractor Thales Alenia Space (Cannes), and including Astrium (Friedrichshafen) responsible for the payload module and for system testing at spacecraft level, Thales Alenia Space (Turin) responsible for the service module, and Astrium (Toulouse) responsible for the telescope, with in excess of a hundred subcontractors.




\bibliographystyle{mnras}
\bibliography{ref.bib} 



\appendix

\section{CO spectra of all clumps}
\label{sec:appendix_A}

Figure~\ref{fig:clumps_lines} shows the spectra of $^{12}$CO, $^{13}$CO and C$^{18}$O (3-2) of all 27 clumps.
For clumps with $^{13}$CO/C$^{18}$O (3-2) intensity higher than 2.4\,K ($3\sigma$ of the main-beam-efficiency corrected CHIMPS data, shown by grey dashed lines), we fit the line profile with a Gaussian function and derive the centre velocity and line width.
Blue dashed lines show the fitted centre velocities of $^{13}$CO and green dashed lines show the fitted centre velocities of C$18$O for each clump. Only for Clump 20 and Clump 22, centre velocities of $^{13}$CO deviate from centre velocities of C$^{18}$O, indicates that $^{13}$CO might be self-absorbed. The evolutionary classes are labeled at the top right.

\begin{figure*}
	\centering
	\includegraphics[width=\textwidth,trim=100 100 100 100,clip]{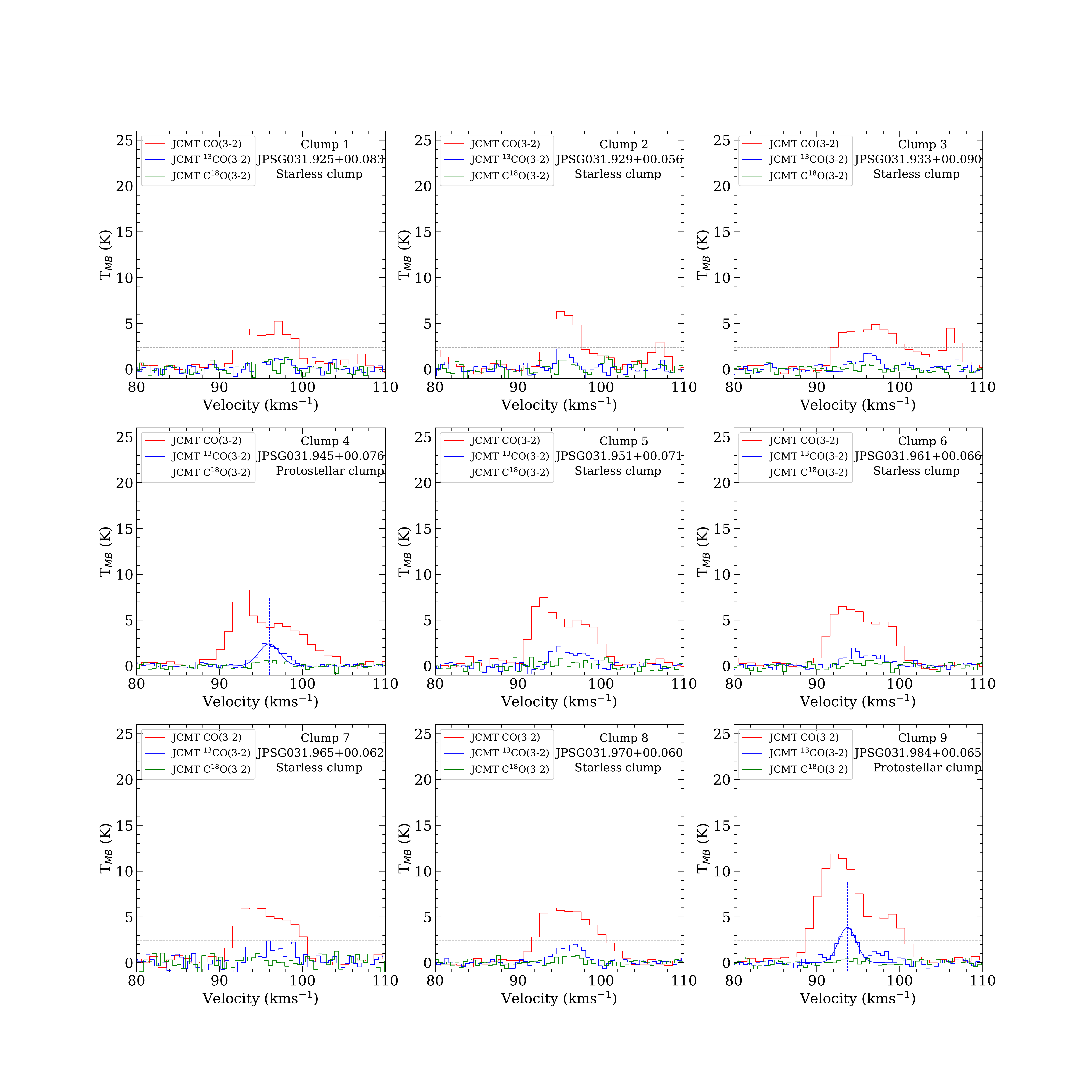}
    \caption{$^{12}$CO, $^{13}$CO and C$^{18}$O (3-2) spectra of each clump. For each plot, the lines of $^{12}$CO, $^{13}$CO and C$^{18}$O (3-2) are coloured red,blue and green, respectively. Grey dashed lines show the $3\sigma$ threshold of main-beam efficiency corrected CHIMPS data ($\sim 2.4\,$K). Blue dashed lines show the fitted centre velocities of $^{13}$CO and green dashed lines show the fitted centre velocities of C$^{18}$O for each clump. The evolutionary classes are labeled at the top-right.}
    \label{fig:clumps_lines}
\end{figure*}

\begin{figure*}
	\centering
	\includegraphics[width=\textwidth,trim=100 100 100 100,clip]{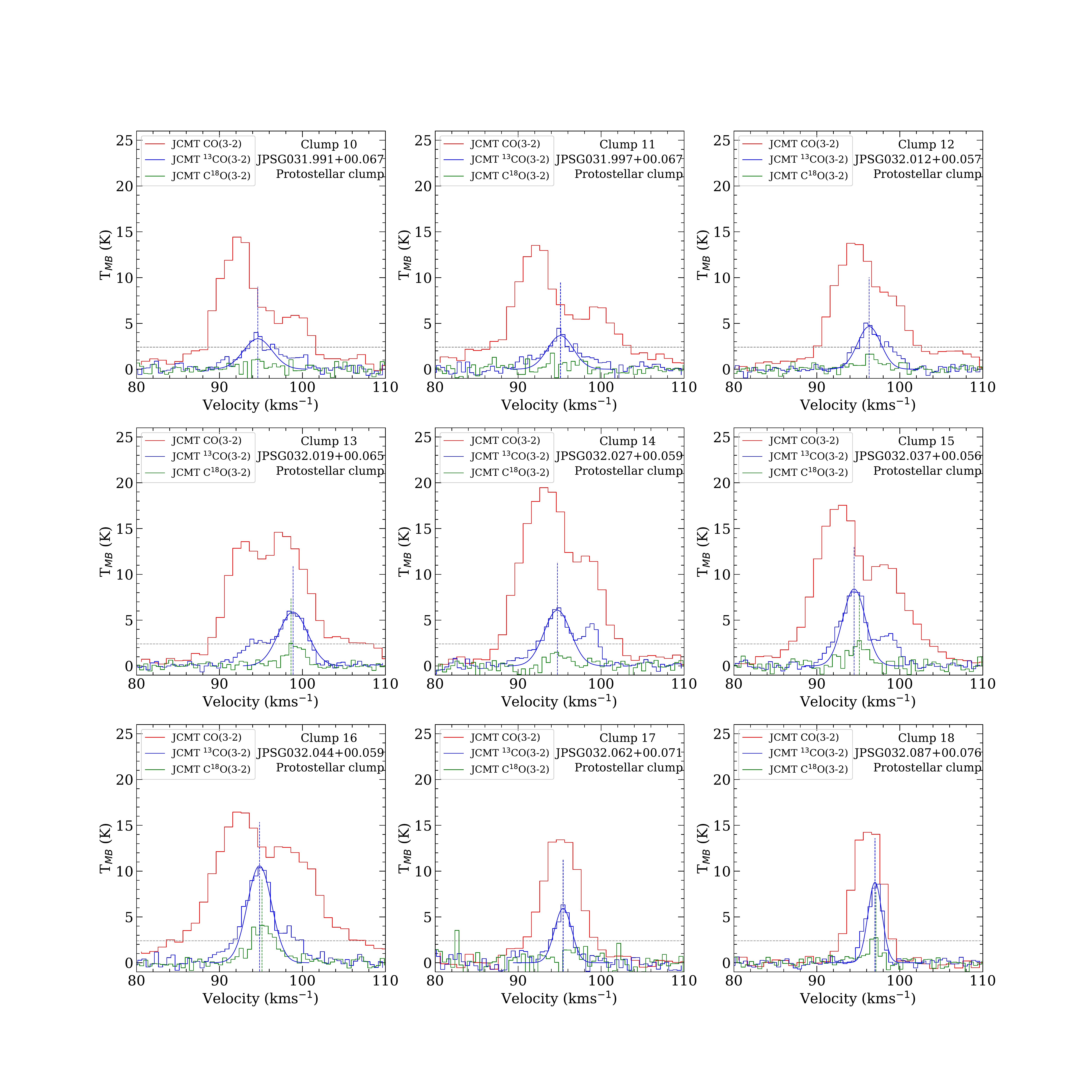}
    \caption{Continued}
\end{figure*}

\begin{figure*}
	\centering
	\includegraphics[width=\textwidth,trim=100 100 100 100,clip]{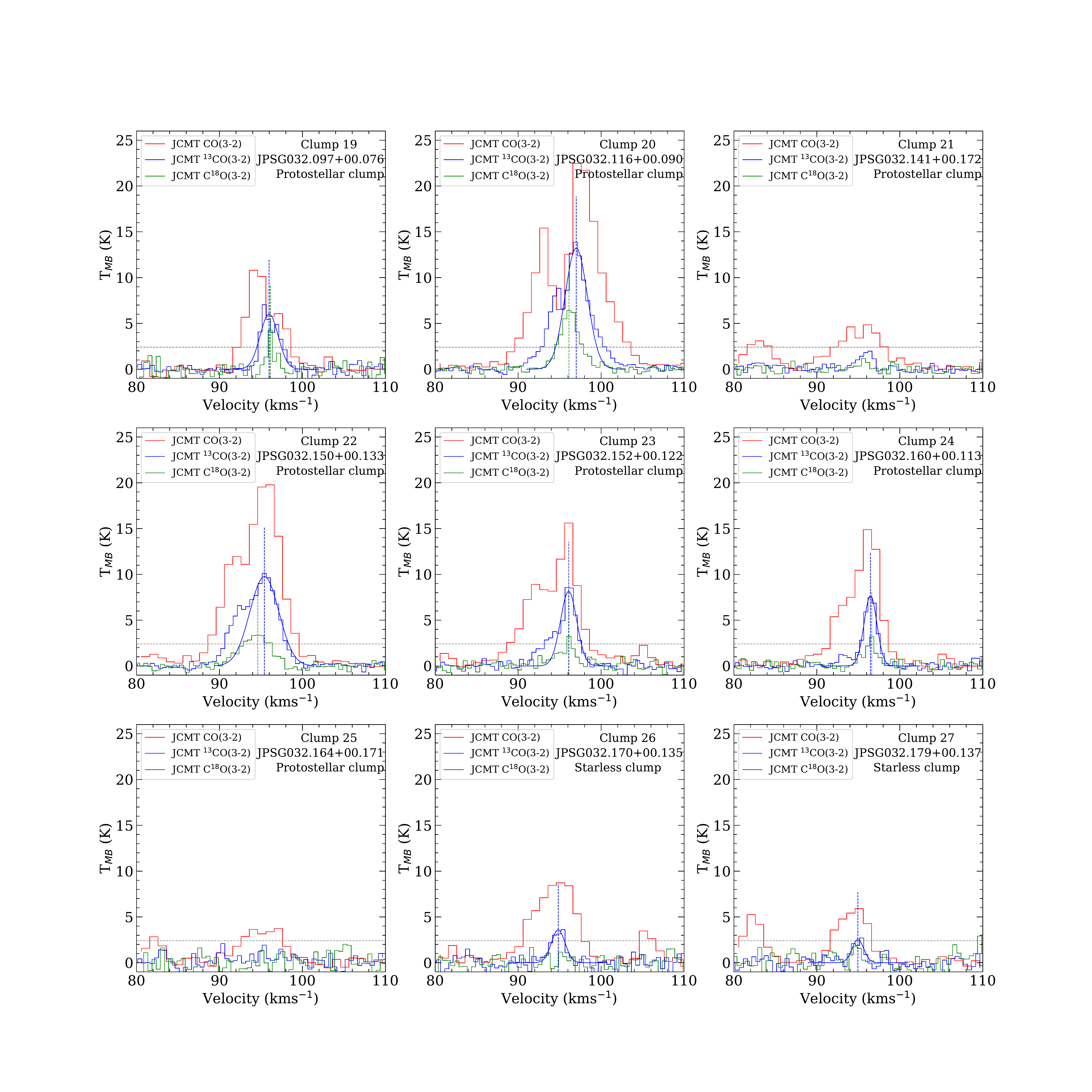}
    \caption{Continued}
\end{figure*}


\bsp	
\label{lastpage}

\end{document}